\documentclass[useAMS,usenatbib]{mn2e}
\usepackage{times}
\usepackage{graphics}
\usepackage{graphicx}
\usepackage{color}
\usepackage{verbatim}
\usepackage{amsmath}
\usepackage{amssymb}
\usepackage{mathtools}
\usepackage{subfigure}
\usepackage{float}
\usepackage{tabularx}
\usepackage{supertabular}
\usepackage{rotating}
\usepackage{longtable}
\usepackage{dpfloat, booktabs}
\usepackage{color}
\usepackage{multicol}
\usepackage{graphicx}
\usepackage{multirow}               
\usepackage{amsfonts}

\usepackage{pifont}

\usepackage{booktabs,caption}
\usepackage[flushleft]{threeparttable}

\usepackage{wrapfig}
\usepackage{subfig}   

\definecolor{Mygrey}{gray}{0.75}

\title[Gas, dust \& stellar population in NGC~1055] {From Stellar Nurseries to Old Stellar Populations: A Multi-wavelength Case of NGC 1055}
\author[S.\ Topal]  {Sel\c{c}uk Topal$^{1}$\thanks{E-mail:
    selcuktopal@yyu.edu.tr}\\
    $^{1}$Van Yuzuncu Yil University, Department of Physics, Van, 65080,
  Turkey\\ }

\begin{document}
\date{Accepted . Received; in original form }
\pagerange{\pageref{firstpage}-\pageref{lastpage}} \pubyear{2024}
\maketitle
\label{firstpage}
%
%
\begin{abstract}
Given the complex nature of galaxies' interstellar medium (ISM), multi-wavelength data are required to probe the interplay among gas, dust, and stellar populations. Spiral galaxies are ideal laboratories for such a goal as they are rich in gas and dust. Using carbon monoxide (CO) along with $GALEX$ far-ultraviolet (FUV) and $Spitzer$ near-infrared (NIR) data we probe the correlations amongst the properties of stellar populations, gas, and dust over the disc of the spiral galaxy NGC~1055 at multiple angular resolutions, i.e. $2$, $4$, and $17$~arcsec corresponding to a linear size of $144$~pc, $288$~pc, and $1.2$~kpc, respectively. Our results indicate an asymmetry in the physical conditions along the galaxy's disc, i.e. the gas is slightly more extended and brighter, and molecular gas mass is higher on the disc's eastern side than the western side. All physical properties (i.e. molecular gas mass, CO line ratios, stellar mass, NIR emission) decrease from the centre going outwards in the disc with some exceptions (i.e. the extinction, FUV radiation, and the [$3.6$]$-$[$4.5$] colour). Our analysis indicates that the colour gets bluer (metallicity increases) halfway through the disc, then redder (metallicity decreases) going outwards further in the disc.

\end{abstract}
\begin{keywords}
  dust, extinction ~- ISM: molecules ~- galaxies: ISM ~- galaxies: spiral
\end{keywords}
%
%
\section{Introduction}
\label{sec:intro}

The Universe hosts many galaxies from different Hubble types with unique evolutionary pathways. Elliptical galaxies have lost almost all of their molecular gas reservoir during their evolution, but spiral galaxies still shine brightly with substantial star formation activity. Spiral galaxies are thus natural laboratories to study recent star formation activities in galaxies. Carbon monoxide (CO) is ubiquitous in the interstellar medium (ISM). It is easily detectable in such cold gas clouds contrary to hydrogen molecule (H$_{2}$), the most abundant molecule of the ISM but extremely difficult to obverse its rotational transitions in such cold environments. CO is, therefore, a useful tracer to probe the gas clouds. 

The ISM is a complex place where many phenomena are simultaneously at play and has multiple phases, i.e. cold/warm neutral components, and warm/hot ionised components, resulting from various heating mechanisms \citep{ro07}. The gas kinetic temperature, density, FUV radiation, shocks, cosmic rays, X-rays, and turbulence can change the chemistry, ionisation level, and structure of the gas, as well as the molecular line ratios in the medium \citep{hol99, mei05, ro07, pan09, ros14}. Multiple tracers are thus required to understand the properties of molecular gas, stellar populations, and dust in galaxies and the evolution of galaxies from blue cloud to red sequence along the galaxy colour-magnitude diagram \citep[e.g.][]{bell04}. 

Molecular gas, dust, and ultraviolet (UV) radiation are closely related. The UV radiation from massive stars has multiple effects on the environment by either suppressing or favouring the star formation. Strong UV radiation could dissociate and ionize cold gas, causing an increase in thermal and radiation pressure in the cloud \citep{kim18}. The UV radiation is also responsible for creating the photon-dominated regions (PDRs) outside the molecular clouds \citep{hol99}, and its overall effect on star formation is negative \citep{mc07}. Molecular mass fraction depends on the interplay among the far-ultraviolet (FUV), interstellar radiation field (ISRF), and dust \citep[e.g.][]{hol71, mark09}. Although the percentage of dust in the ISM is small compared to the other constituents, it has a vital role in shaping the structure and chemistry of the medium. Dust grain surfaces act as catalysts for various chemical reactions to produce molecules, such as H$_{2}$, the fuel for star formation \citep[e.g.][]{hol71,wak17}. 

Integrated CO intensity and line ratios show radial variations depending on the position over the galaxy disc, i.e. intensities, velocity dispersion, and surface density decrease from the centre to the outskirts and are positively correlated with the local specific star formation rate (sSFR) surface density \citep[e.g.][]{topal20, brok21, ler22, koda23}. Isotopic ratios, such as R$_{11}$, provide us with some insights on the column density and opacity of the gas, i.e. higher R$_{11}$ ratio indicates thinner gas, allowing us to see deeper into the gas cloud \citep[e.g.][]{pine08}. The ratio of R$_{11}$ could also give information on the possible mechanical feedbacks in the medium, i.e. stellar winds or supernova explosions could make the gas more diffuse in some regions. Such mechanical feedback could cease the star formation in the gas clouds while triggering star formation by making the gas denser in other parts of the clouds \citep[e.g.][]{wha08,naga09,iff15}. The ratios of R$_{21}$ and R$_{31}$ provide insights into the average temperature and density, as higher transitions of the same species require an environment with higher temperature and density. 

Younger, hotter, and more massive stars are the source of FUV radiation in the ISM, which could heat the surrounding dust and make it shine brightly at $3.6\,\micron$ near-infrared (NIR) emission \citep{meit12}, and change the chemistry in the medium by causing transitions between atomic and molecular species, such as the H$\small{\rm I}$-to-H$_{2}$ transition \citep[e.g.][]{sten14,bia17, bia20}. The $3.6\,\micron$ emission is almost free of extinction. Old stars and young massive stars or a combination of both could result in making the at $3.6\,\micron$ emission brighter. However, the lights from old stars could also contain polycyclic aromatic hydrocarbons (PAHs) up to the level of $15\%$ \citep{mei12}. \citet{que15} studied the total light at $3.6\,\micron$ in galaxies from the $Spitzer$ Survey of Stellar Structure in Galaxies (S$^{4}$G). The survey includes spirals, ellipticals, and also dwarf galaxies. They conclude that up to $10\%$\,-\,$30\%$ of the emission at $3.6\,\micron$ has dust origin and is related to star formation. The $3.6\,\micron$ emission provides us with information on old stellar populations and the effects of UV radiation in the ISM. 

Stellar populations with old K and M giants are expected to have $-0.2<$~[$3.6$]$-$[$4.5$]~$<0$ in colour because of the strong CO absorption at $4.5\,\micron$ causing a decrease in $4.5\,\micron$ flux. However, young stars have redder [$3.6$]$-$[$4.5$] colour because of less deeper CO absoprtion \citep{pel12, mei14}. The colour is almost always positive in the case of non-stellar origin \citep{que15}. The range for the colour for diffuse dust is given as $0.2$\,$<$\,[$3.6$]$-$[$4.5$]\,$<$\,$0.7$ \citep{wil04,pah04,pel12,que15}. \citet{pel12} probed the [$3.6$]$-$[$4.5$] colour gradient in early-type galaxies (ETGs), and they found that the galaxies show redder colour and a decrease in metallicity with galactocentric radius. The $[3.6]$$-$$[4.5]$ colour also shows dependence on redshifts \citep{lab13, smit14}. In general, the colour [$3.6$]$-$[$4.5$]$\,>$\,$0$ indicates a non-stellar origin, i.e. dust heated by the ISRF, while [$3.6$]$-$[$4.5$]$\,<\,0$ indicates a stellar origin, i.e. the atmospheres of old stars. 

There is a bimodality in galaxies in terms of many properties, such as colour, spectral class, metallicity, morphology, internal properties, and efficiency in creating stars \citep[e.g.][]{mad02, kauff03, bald04, bri04, fab07}. Given the timescale of galaxy evolution, it is impossible to observe the evolution of a single galaxy throughout its lifetime. Observing multiple galaxies at different redshifts with similar linear resolution and sensitivity is also challenging. It is therefore essential to study a galaxy with multiple tracers obtained at the same resolution and different positions in the galaxy, as we do in the current study.

In this study, we gathered data for NGC~1055 to probe the correlations among molecules, dust, and stars in a gaseous disc of a spiral galaxy. NGC~1055 is a nearby edge-on barred spiral galaxy with a morphological type of SBb (NASA Extragalactic Database, NED), and located at $14.8$~Mpc distance \citep{wright06}, equivalently an angular scale of $\approx72$~pc~arcsec$^{-1}$. NGC~1055 is fully mapped in CO, FUV, $3.6$, and $4.5$ microns. The angular extent, brightness, and number of emission features in the edge-on disc of NGC 1055 make it an excellent target for resolving many unknowns about stellar evolution and its effects on dust and molecular clouds in a spiral galaxy. For the first time for this galaxy, our goals are; (1) to study star-forming gas in the centre and throughout the disc of the galaxy using the combination of CO, FUV, and NIR data; (2) to probe correlations among the physical parameters, such as FUV and NIR fluxes, the mass of molecular and stellar components, between mass and $[3.6]$$-$$[4.5]$ colour, and also between each physical property and the galactocentric radius; (3) to probe the effects of the stellar populations on the physics of the ISM; and finally, (4) to probe the small and large structures, so the effects of angular resolution on the results.

The upcoming sections are organised as follows. We present the literature data and the analysis in Section~\ref{sec:datan}. Our discussion on the results and our final remarks are in Sections 3 and 4, respectively.


\section{Data and analysis}
\label{sec:datan}

\subsection{Archival CO Data}
\label{sec:co}
The data for $^{12}$CO(1--0) and its isotopologue $^{13}$CO(1--0) were taken from the CO Multi-line Imaging of Nearby Galaxies (COMING) survey conducted using the Nobeyama $45$\,m single-dish telescope \citep{sor19}. The Nobeyama has a $17$~arcsec beam size at the observed frequency of $115$~GHz. This corresponds to a linear diameter of about $1.2$ kpc over the galaxy. he CO data cubes have a velocity resolution of $10$~km~s$^{-1}$. The root-mean-square (RMS) noise in the cubes was estimated to be $70$~mK and $32$~mK for $^{12}$CO(1--0) and $^{13}$CO(1--0), respectively.

\begin{figure*}
\begin{center}
  \includegraphics[width=8cm,clip=]{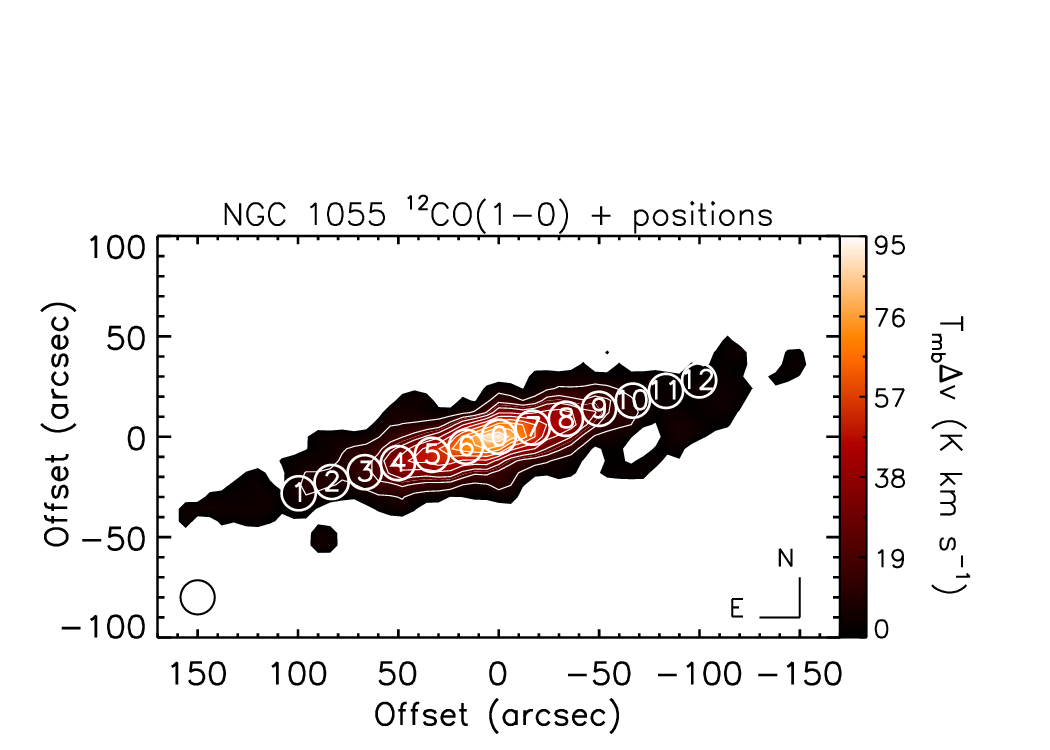} 
  \includegraphics[width=8cm,clip=]{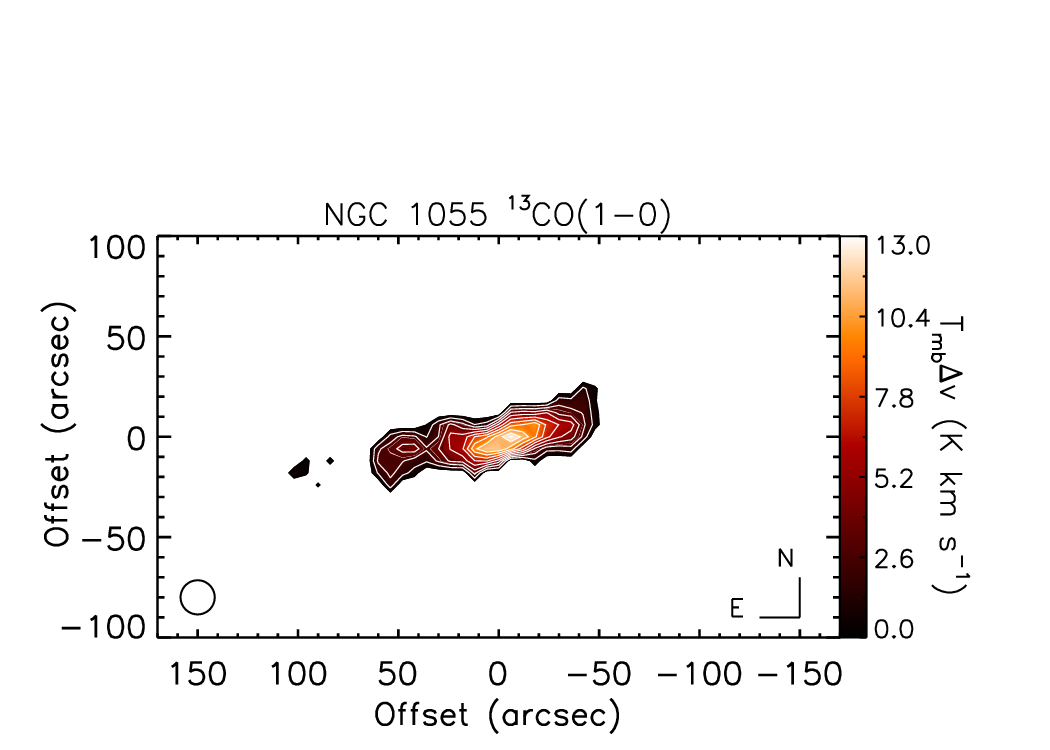} \\
      \includegraphics[width=8cm,clip=]{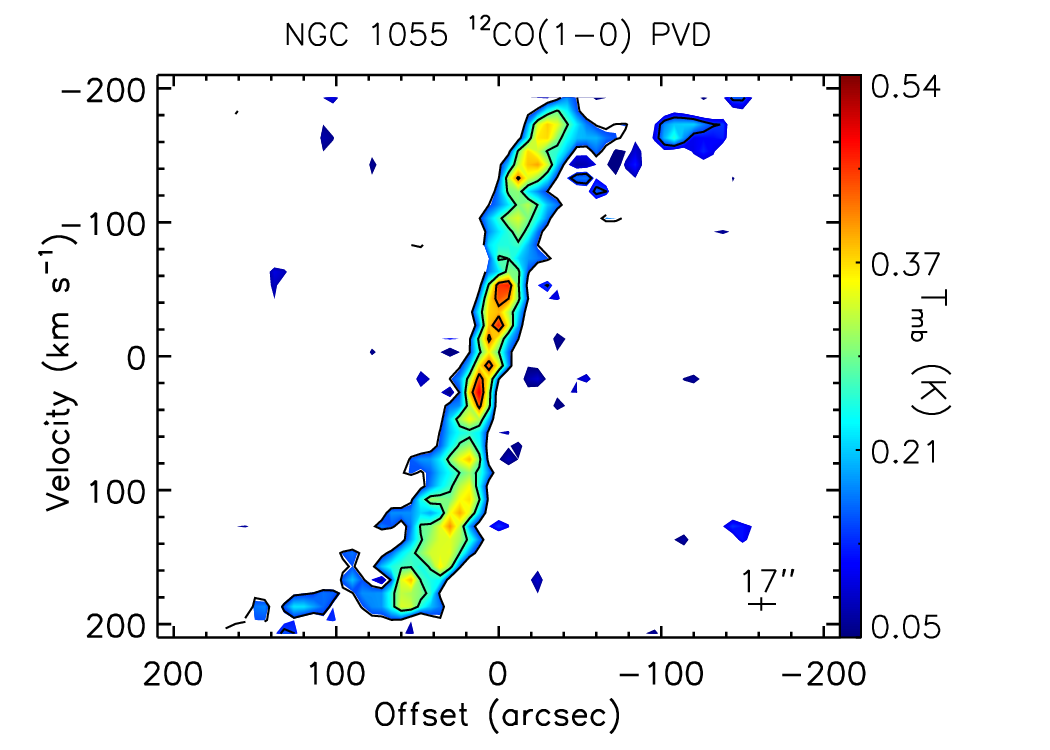} 
   \includegraphics[width=8cm,clip=]{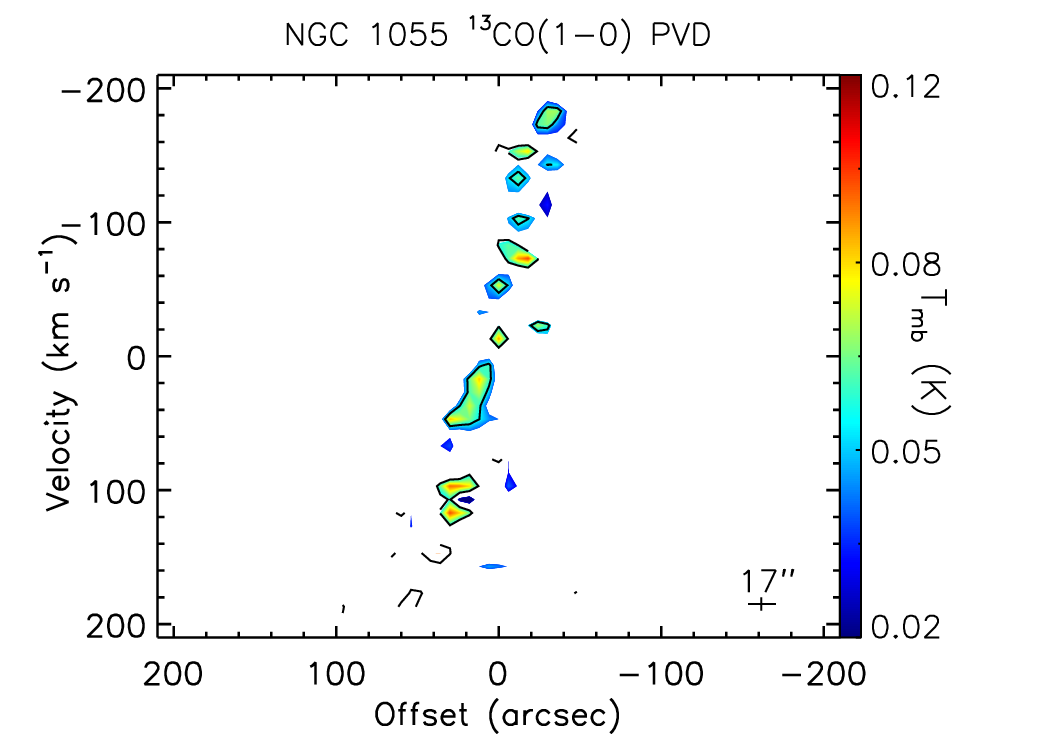}\\
  \caption{Top: CO moment 0 maps are shown. Integrated intensity contours shown by white lines are spaced by 10\% of the peak intensity of $95$ and $13$~K~km~s$^{-1}$ for $^{12}$CO(1--0) and $^{13}$CO(1--0), respectively. $1\sigma$ RMS in the moment map is $6.4$~K~km~s$^{-1}$ for $^{12}$CO and $0.8$~K~km~s$^{-1}$ for $^{13}$CO. The positions over the $^{12}$CO(1--0) moment map are represented by white circles. The diameter of each circle is $17$ arcsec or $1.2$ kpc over the galaxy. The numbers inside each white circle are as listed in Table~\ref{tab:ratios}. The direction in the sky is shown in the panels. Bottom: The PVDs for both lines are shown. The PVDs have contour levels spaced by $3\sigma$, where $1\sigma$ RMS noise on the PVDs are $50$~mK and $22$~mK for $^{12}$CO and $^{13}$CO, respectively. The angular resolution (horizontal line) and velocity resolution (smaller vertical line) are shown in each panel. }
    \label{fig:pos}
\end{center}
\end{figure*}

The CO intensity maps (hereafter moment 0) and position velocity diagrams (PVDs) for both CO lines were produced after defining the extent of the CO emission across the galaxy. To do that we applied Gaussian smoothing and clipping at the $3~\times~$RMS noise to the data cubes. We obtained CO PVDs by taking a three-pixel slice in the vertical direction through the major axis (see \citealt{topal16} for more details on the method). The maps are shown in Figure~\ref{fig:pos}.

%

\subsection{CO intensities, gas mass, and line ratios}
\label{sec:intenmass}

\subsubsection{Line intensities and total molecular gas mass}
\label{sec:linemass}
Including the centre, we selected $13$ positions, each with a diameter of $17$ arcsec, side by side along a linear cut, crossing the galactic centre, to extract the beam-averaged CO spectra (see Fig.~\ref{fig:pos}, and Figures~\ref{fig:pro} and ~\ref{fig:profit} in the Appendix). The spectra were extracted from the (masked) data cubes using MIRIAD task \emph{imspec}. At two positions beyond $50$ arcsec on the northwestern side (NW) of the disc and one position on the southeastern (SE), there is no $^{13}$CO(1--0) detected, i.e. the positions $3$, $9$, and $10$ (see Figure~\ref{fig:pos}). Furthermore, no $^{13}$CO(1--0) emission is found at the two farthest positions on the NW side, i.e. the positions $11$ and $12$ (see Figure~\ref{fig:pos}), possibly as a result of our masking procedure (see Section~\ref{sec:co}). We defined an upper limit integrated $^{13}$CO(1--0) intensity at those positions as $3\times\sigma_{\rm RMS}\times \delta v\times\sqrt{N}$, where $\sigma_{\rm RMS}$ is the spectral noise (i.e. the rms on the $^{12}$CO(1--0) spectrum at those positions), $\delta v\,=\,10$~km~s$^{-1}$ for the CO data set, and $N$ represents the number of channels covered by the line emission.

\begin{figure*}
\begin{center}
  \includegraphics[width=8.0cm,clip=]{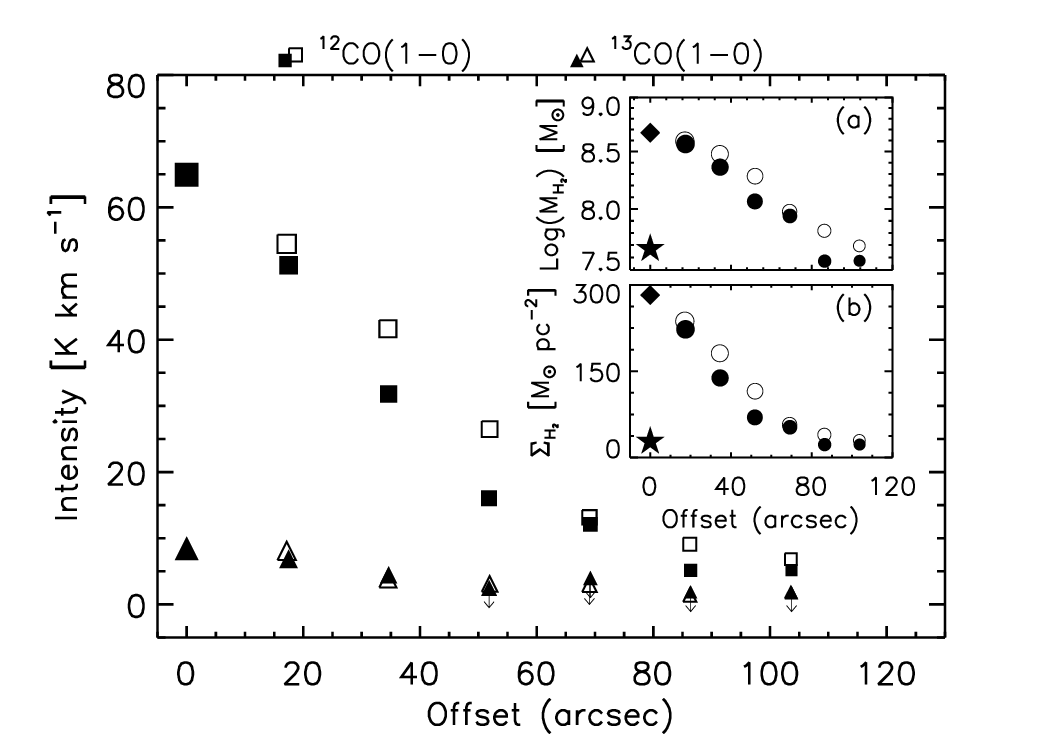}
   \includegraphics[width=8.0cm,clip=]{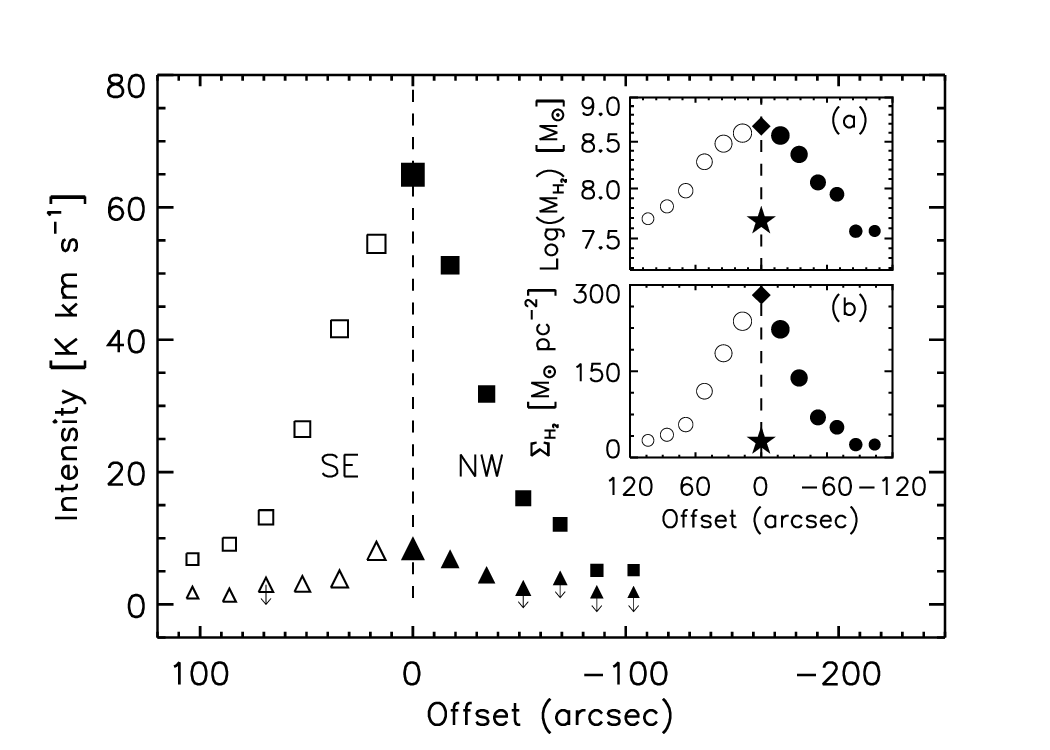}\\
  \caption{Left: The CO intensity, $\int T_{\rm mb}$ d$v$, $M_{{\rm H}_2}$, and $\Sigma_{\rm H_{2}}$ are shown. Open symbols represent the positions in the southeastern (SE), while filled symbols are for the positions in the northwestern (NW). The values for the $M_{{\rm H}_2}$ and $\Sigma_{\rm H_{2}}$ are shown in the embedded panels $a$ and $b$. Right: Same as the left image but the change in $\int T_{\rm mb}$ d$v$,  $M_{{\rm H}_2}$, and $\Sigma_{\rm H_{2}}$ is shown on both sides of the disc separately. The dashed black lines in the image represent the central position in the disc. The symbols at zero offset represent the values at the central position. The diamond symbols in the embedded panels also represent the $M_{{\rm H}_2}$ and $\Sigma_{\rm H_{2}}$ values at the central position (as the star symbols), but calculated using the standard $X_{\rm CO}$ value representative of the Milky Way disc, i.e. $X_{\rm CO}$\,=\,$2\times10^{20}$~[cm$^{-2}$~(K~km~s$^{-1}$)$^{-1}$] (see Section~\ref{sec:linemass}).}
    \label{fig:massinten}
\end{center}
\end{figure*}

We performed the fitting procedure using IDL code MPFIT \citep{ma09} to calculate the CO intensities as $\int T_{\rm mb}$ d$v$ [K~km~s$^{-1}$], where $T_{\rm mb}$ is the antenna main beam brightness temperature. The best-fit parameters were defined based on the smallest $|1-\chi_{\rm r}^2|$, where the $\chi_{\rm r}^2$ is the reduced $\chi^2$ as $\chi_{\rm r}^2$\,=\,$\chi^2$ / $DOF$, and $DOF$ is the degrees of freedom. CO line intensities obtained at the selected positions are shown in Figure~\ref{fig:massinten} and listed in Table~\ref{tab:ratios}.

%
%

$^{12}$CO(1--0) is a well-known tracer to estimate the total H$_{2}$ mass in galaxies \citep[e.g.][]{young91}. We calculated the beam-averaged molecular gas mass, $M_{{\rm H}_2}$\,[$M_{\odot}$], and molecular gas surface density, $\Sigma_{\rm H_{2}}$\,[M$_{\odot}$~pc$^{-2}$], at the $13$ positions. To estimate the $M_{{\rm H}_2}$ and $\Sigma_{\rm H_{2}}$, the factor to convert the CO intensity to the $M_{{\rm H}_2}$ (i.e. $X_{\rm CO}$ [cm$^{-2}$~(K~km~s$^{-1}$)$^{-1}$]) should be known. However, defining $X_{\rm CO}$ is not trivial as it is subject to environmental characteristics indicated by both observations and models \citep[e.g.][]{accu17,gong20}. The variation of $X_{\rm CO}$ depends on the method applied, metallicity, and resolution of the data \citep{bol13}, all have the potential to cause additional uncertainty in the $X_{\rm CO}$. 

Many literature studies suggest that the value of $X_{\rm CO}$\,=\,$2\times10^{20}$ is suitable for nearby galaxies \citep{bol13}, although it could be ten times lower at the central kpc \citep[e.g.][]{bo08, sand13}. \citet{isra20} studied the central molecular zones in many galaxies, including NGC~1055, with an angular resolution of $22$~arcsec, and found an average value of $X_{\rm CO}$\,=\,$0.2\times10^{20}$ for the central kpc of the galaxies in their sample, which is ten times lower than the standard value for the Milky Way disc, i.e. $X_{\rm CO}$\,=\,$2\times10^{20}$~[cm$^{-2}$~(K~km~s$^{-1}$)$^{-1}$]. It is, therefore, more viable to use two different $X_{\rm CO}$ values for the centre and the disc of NGC~1055. We adopted a value of $0.2\times10^{20}$ and $2\times10^{20}$ for $X_{\rm CO}$ for the centre and the disc of NGC~1055, respectively. The corresponding values of $\alpha_{\rm CO}\,\equiv\,\Sigma_{\rm H_{2}}$ / $I_{\rm 1-0}$ [$M_{\odot}$ (K~km~s$^{-1}$~pc$^{2}$)$^{-1}$] are $\alpha_{\rm CO}$\,=\,$0.43$ and $\alpha_{\rm CO}$\,=\,$4.35$ for the centre and the disc, respectively (a factor of $1.36$ was applied to account for He). In addition, the $M_{{\rm H}_2}$ and $\Sigma_{\rm H_{2}}$ values were calculated at the central position using the $X_{\rm CO}$ value representative of the Milky Way disc, and are presented in the associated figures in the following sections (see Section~\ref{sec:reso17}). We used the standard expression to calculate the $M_{{\rm H}_2}$ at all positions studied \citep{bol13}. 

The luminosity distance, D$_{\rm L}$, necessary for the mass calculation was estimated using the cosmology calculator \citep{wright06} for H$_{0}$\,=\,$67.4$~kms$^{-1}$Mpc$^{-1}$, $\Omega_{\rm m}$\,=\,$0.3157$, and $\Omega_{\Lambda}$\,=\,$0.6843$ \citep{planck20}, and $z\,=\,0.00332$ (NED) for a flat universe. The variation of $M_{{\rm H}_2}$ and $\Sigma_{\rm H_{2}}$ are shown in the embedded panels of Figure~\ref{fig:massinten}.

\begin{figure}
\begin{center}
  \includegraphics[width=8.0cm,clip=]{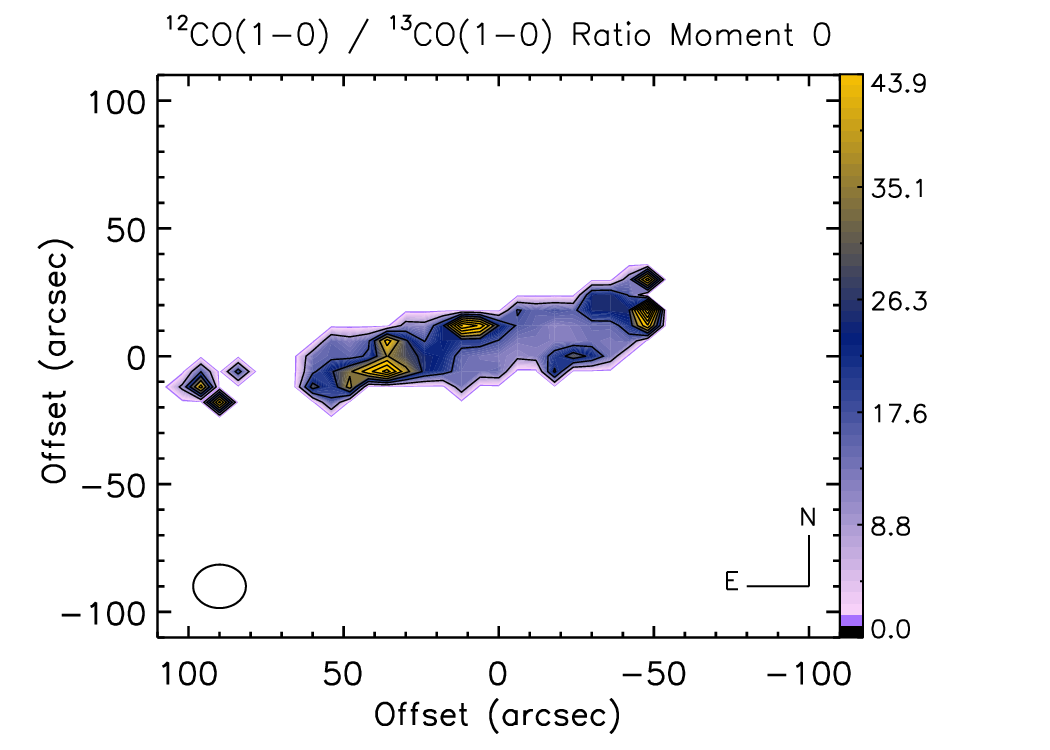} \\
  \caption{The ratio of the moment 0 maps is shown. Contour levels are from  $\%10$ to  $\%100$ of the peak ratio in the $\%10$ percent step. The beam and the direction in the sky are also shown in the panel.}
    \label{fig:ratmap}
\end{center}
\end{figure}

\subsubsection{Line ratios}
\label{sec:lratio}
\begin{figure*}
\begin{center}
  \includegraphics[width=8cm,clip=]{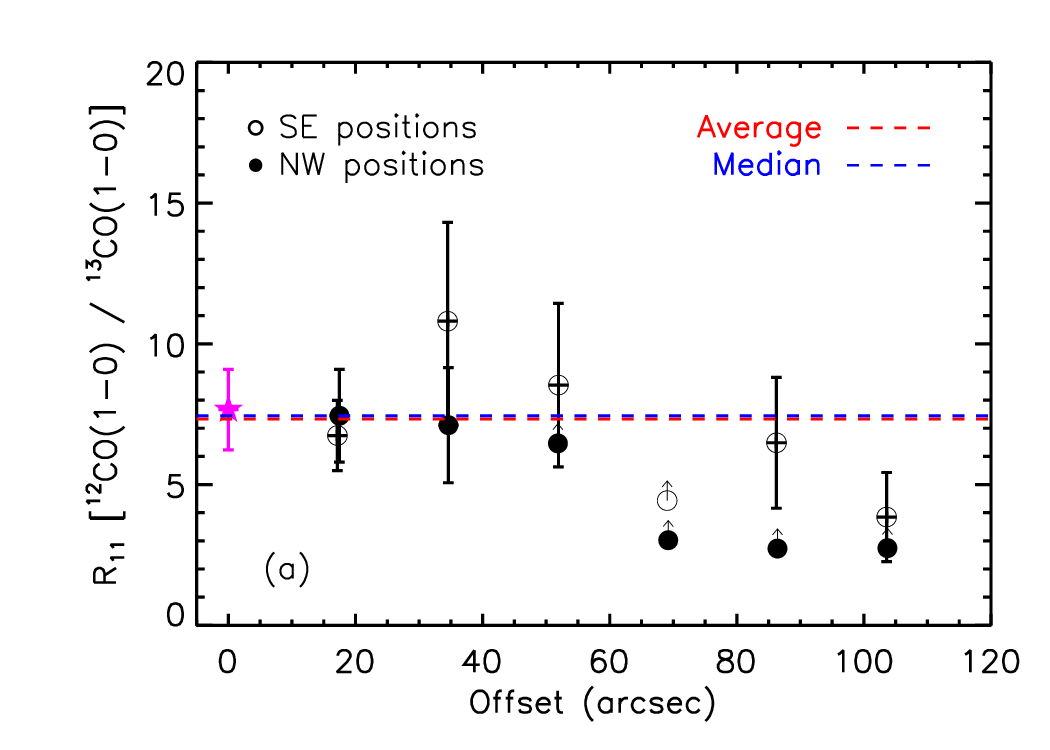}
   \includegraphics[width=8cm,clip=]{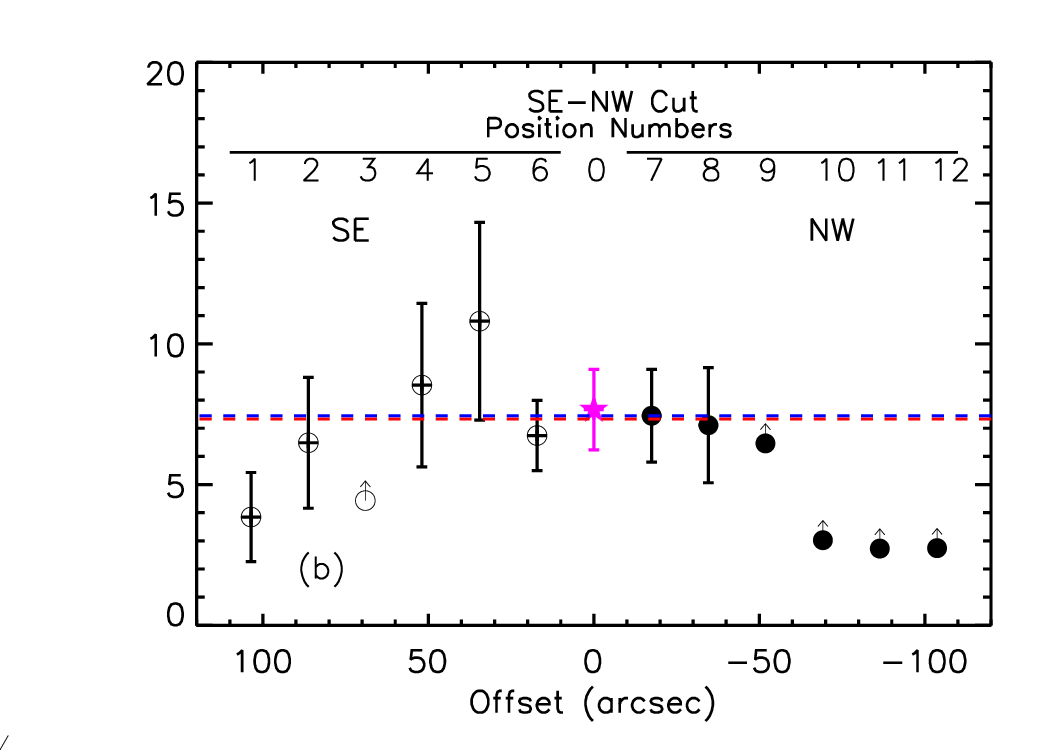}
  \caption{The R$_{11}$ ratios are shown as a function of the offsets over the disc. The filled and open black circles show the ratios in the discs' NW and SE, respectively, while the magenta star symbol in both panels indicates the ratio at the central position. The position numbers are also shown in the right panel. The values shown with arrows are lower limits.}
    \label{fig:lineratios}
\end{center}
\end{figure*}

The ratio of $^{12}$CO(1--0) to $^{13}$CO(1--0) (hereafter R$_{11}$) along the galaxy's major axis is listed in Table~\ref{tab:ratios} and also shown in Figure~\ref{fig:lineratios}. The  R$_{11}$ ratios at position $3$ and positions $9-12$ are lower limits. We also acquired the literature data for the ratio of $^{12}$CO(2--1) / $^{12}$CO(1--0) (hereafter R$_{21}$) and $^{12}$CO(3--2) / $^{12}$CO(1--0) (hereafter R$_{31}$) in the central region of the galaxy observed at a similar aperture, i.e. $22$~arcsec \citep{isra20, mao10}. 

%
%

%
\begin{table*}
\begin{threeparttable}
\centering
  \caption{Archival data and their usage}
  \begin{tabular}{llccl}
    \hline
    Observations&Molecule/Tracer& RMS noise$^{\rm *}$&Resolution [arcsec]& Purpose \\
    \hline 
    \multirow{2}{*}{COMING}&$^{12}$CO(1--0)&$70$~mK& \multirow{2}{*}{$17$}& \multirow{2}{*}{Moment maps, PVDs, changes in R$_{11}$, $M_{{\rm H}_2}$, and $\Sigma_{\rm H_{2}}$}\\ 
    &$^{13}$CO(1--0)&$32$~mK&&\\ \\
    
        \multirow{2}{*}{GALEX}&\multirow{2}{*}{FUV $154$~nm}&\multirow{2}{*}{$5\times10^{-5}$~Jy}&$17$&Changes in FUV and correlations with $M_{{\rm H}_2}$ and R$_{11}$\\
                       &&&$4$&Changes in FUV and F$_{3.6\micron}$/F$_{FUV}$ ratio, that is, the extinction\\  \\
    
        \multirow{3}{*}{SPITZER}&\multirow{2}{*}{NIR $3.6~\micron$}&\multirow{2}{*}{0.0072 MJy/sr}&$17$&Changes in colour, NIR, and correlations with $M_{{\rm H}_2}$ and R$_{11}$ \\
    &\multirow{2}{*}{NIR $4.5~\micron$}&\multirow{2}{*}{0.0093 MJy/sr}&$4$&Changes in colour, NIR, FUV, $M_{\star}$, the extinction, and correlations with FUV and $M_{\star}$\\            
    &&&$2$&Changes in colour, NIR and $M_{\star}$\\\\
    
    \hline
  \end{tabular}		
  \label{tab:litedat}
  \begin{tablenotes}
  \item $^{\rm *}$ The RMS noise is estimated for the original angular resolution, that is, $2$, $4$, and $17$~arcsec for NIR, FUV, and CO data, respectively. The noise for the CO data was estimated in this study. The noise listed for the NIR and FUV data was taken from \citet{she10} and \citet{gil07}, respectively.
  \end{tablenotes}
  \end{threeparttable}
\end{table*}
%

%
%
%
\begin{table*}
\begin{threeparttable}
\centering
  \caption{The $^{12}$CO(1--0) and $^{13}$CO(1--0) intensities and their ratios along the disc of NGC~1055.}
  \begin{tabular}{cccccc}
    \hline
    Position&$^{12}$CO(1--0)~[K~km~s$^{-1}$]&(S/N)&$^{13}$CO(1--0)~[K~km~s$^{-1}$]&(S/N)&$^{12}$CO(1--0) / $^{13}$CO(1--0)\\
    \hline 
    0&$64.93\,\pm\,3.58$ &$(12)$&$8.47\,\pm\,1.51$ &$(4)$&$\phantom{0}7.66\,\pm\,1.42$\\ 
    1&$\phantom{0}6.84\,\pm\,1.86$ &$(4)$&$1.78\,\pm\,0.55$ &$(3)$&$\phantom{0}3.85\,\pm\,1.58$\\ 
    2&$\phantom{0}9.09\,\pm\,1.61$ &$(5)$&$1.40\,\pm\,0.44$ &$(3)$&$\phantom{0}6.48\,\pm\,2.33$\\ 
    3&$13.16\,\pm\,1.93$ &$(7)$&$\phantom{0000}\le2.97$&&$\phantom{00000}\ge4.43$\\ 
    4&$26.50\,\pm\,2.66$ &$(11)$&$3.10\,\pm\,1.01$ &$(4)$&$\phantom{0}8.53\,\pm\,2.91$\\ 
    5&$41.67\,\pm\,2.69$ &$(15)$&$3.86\,\pm\,1.23$ &$(6)$&$10.80\,\pm\,3.51$\\ 
    6&$54.48\,\pm\,3.57$ &$(10)$&$8.08\,\pm\,1.40$ &$(5)$&$\phantom{0}6.74\,\pm\,1.25$\\ 
    7&$51.24\,\pm\,3.41$ &$(12)$&$6.88\,\pm\,1.45$ &$(5)$&$\phantom{0}7.44\,\pm\,1.65$\\ 
    8&$31.80\,\pm\,2.82$ &$(14)$&$4.47\,\pm\,1.22$ &$(4)$&$\phantom{0}7.11\,\pm\,2.05$\\ 
    9&$16.03\,\pm\,2.14$ &$(8)$&$\phantom{0000}\le2.48$&&$\phantom{00000}\ge6.46$\\ 
    10&$12.08\,\pm\,2.05$ &$(6)$&$\phantom{0000}\le4.00$&&$\phantom{00000}\ge3.02$\\ 
    11&$\phantom{0}5.15\,\pm\,1.56$ &$(4)$&$\phantom{0000}\le1.89$&&$\phantom{00000}\ge2.72$\\ 
    12&$\phantom{0}5.18\,\pm\,1.39$ &$(5)$&$\phantom{0000}\le1.89$&&$\phantom{00000}\ge2.74$\\             
    \hline
  \end{tabular}		
  \label{tab:ratios}
    \begin{tablenotes}
  \item \textit{Notes.} The S/N values presented in parentheses represent the peak signal-to-noise ratio in the spectra. The RMS noise in the spectra is estimated by considering only the emission-free velocity channels (see Figure~\ref{fig:profit}). The integrated $^{13}$CO(1--0) intensities at positions $3$ and $9$-$12$ are upper limits; consequently, the associated ratios are lower limits.
  \end{tablenotes}
  \end{threeparttable}
\end{table*}

We look into the ratios of the integrated line intensity maps of $^{12}$CO(1--0)  and $^{13}$CO(1--0), providing us with a more general picture of any variations in the R$_{11}$ ratios across the entire gaseous disc of the galaxy. As both CO data cubes are identical, i.e. having the same angular resolution, pixel size, and sky coverage, we directly took the ratio of the CO maps (Figure~\ref{fig:ratmap}).
%

\begin{figure*}
\begin{center}
  \includegraphics[width=17cm,clip=]{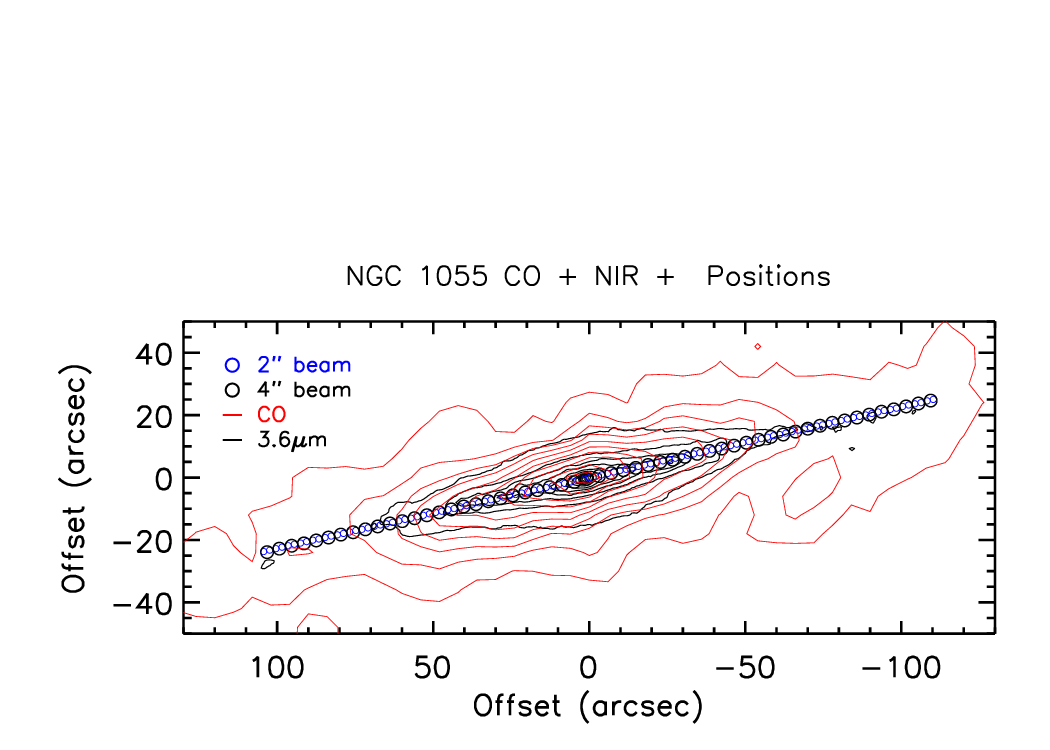}\\
  \caption{The selected positions at an angular resolution of $2$~arcsec (i.e. blue circles, $110$ positions in total) and $4$~arcsec (i.e. black circles, $55$ positions in total) are overlaid on galaxy's CO (red) and $3.6\,\micron$ (black) contour maps.}
    \label{fig:gasdustpos}
\end{center}
\end{figure*}

\subsection{Far-ultraviolet (FUV) and near-infrared (NIR) data}
\label{sec:farnear}
To investigate potential correlations amongst molecular gas properties, dust, and stellar populations, data from the literature at FUV and NIR wavelengths across the disc of NGC 1055 were also collected. FUV data were taken from the Galaxy Evolution Explorer survey (GALEX, \citealt{gil07}) while NIR data at $3.6\,\micron$ and $4.5\,\micron$ (hereafter F$_{FUV}$, F$_{3.6\micron}$ and F$_{4.5\micron}$, respectively) were taken from \emph{Spitzer Space Telescope} survey the Infrared Array Camera data archive (\citealt{faz04,wer04}). NIR and FUV maps have a resolution of about $2$ and $4$~arcsec or a linear size of $144$~pc and $288$~pc over the galaxy, respectively. After applying the necessary unit conversion to obtain the GALEX and \emph{Spitzer} data in Jy, the FUV and NIR fluxes were smoothed with a Gaussian kernel with an FWHM of $17$~arcsec to compare them with the CO observations. Please see \citet{topal20} for more details on the method. 

Along and within the same linear cut where the CO intensities were extracted (see Section~\ref{sec:linemass}), F$_{3.6\micron}$ and F$_{4.5\micron}$ flux densities were obtained at three angular resolutions (i.e. $2$, $4$ and $17$~arcsec), whereas the F$_{FUV}$ fluxes were obtained at two (i.e. $4$ and $17$~arcsec). First, we calculated the F$_{3.6\micron}$ and F$_{4.5\micron}$ fluxes at $17$~arcsec, to probe the correlation between the NIR flux and molecular gas-related properties such as $M_{{\rm H}_2}$ and CO line ratios (a total of $13$ positions). Second, we calculated the F$_{3.6\micron}$ and F$_{4.5\micron}$ flux densities at the angular resolution of FUV data, i.e. $4$~arcsec, to compare the colour with the FUV flux densities (a total of $55$ positions). Lastly, we obtained the NIR flux densities at their original resolution of $2$~arcsec to probe the change in the colour and stellar mass across the disc (a total of $110$ positions, see Section~\ref{sec:colmass}). All selected positions at an angular resolution of $2$~arcsec and $4$~arcsec are shown over the galaxy in Figure~\ref{fig:gasdustpos}. 

\begin{figure*}
\centering
      \includegraphics[width=15cm,clip=]{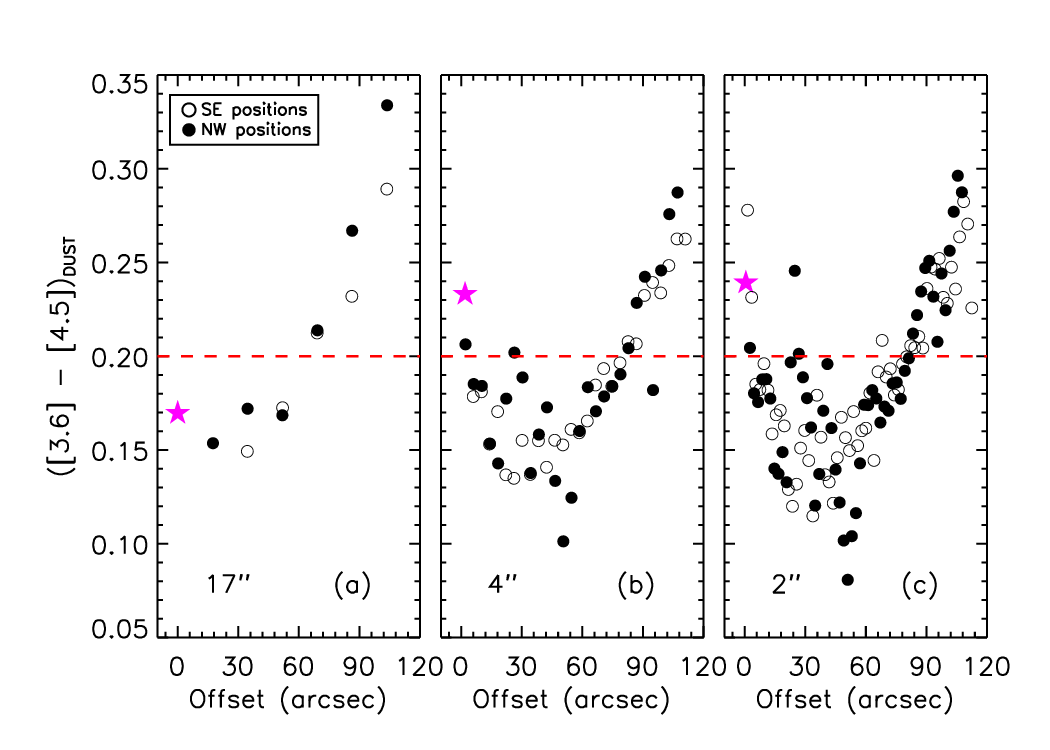} \\
  \caption{The variation of the dust colour with radius is shown for three angular resolutions, i.e. $17$, $4$, and $2$~arcsec, respectively. The horizontal dashed red line in the panels represents the lowest end of the [$3.6$]$-$[$4.5$] colour range for diffuse dust \citep{que15}. The magenta star indicates the value at the central position.}
    \label{fig:colres}
\end{figure*}

Similarly, the values of F$_{FUV}$ were calculated at a resolution of $4$ arcsec to study the change in the F$_{FUV}$ across the galaxy's major axis, and at the resolution of the CO data to probe the correlation between the F$_{FUV}$ and the molecular gas-related properties. All the parameters obtained at different angular resolutions are presented and discussed in Section~\ref{sec:resdis}.

\subsection{\emph{Spitzer} [$3.6$]$-$[$4.5$] colour and stellar mass}
\label{sec:colmass}
We estimated the [$3.6$]$-$[$4.5$] colour as follows. We calculated the apparent magnitudes of $3.6\,\micron$ and $4.5\,\micron$ at each position using the standard expression of $F_{v}\,=\,F_{0}\,\times\,10^{-m/2.5}$, where F$_{0}$ is the zero-point flux densities of $280.9$~Jy and $179.7$~Jy for $3.6\,\micron$ and $4.5\,\micron$, respectively \citep{re05}. F$_{v}$ is the F$_{3.6\micron}$ and F$_{4.5\micron}$ flux densities (in Jy) estimated at each position and for three different angular resolutions (see Section~\ref{sec:farnear}), and $m$ is the apparent magnitudes at $3.6\,\micron$ and $4.5\,\micron$. Using the apparent magnitudes at both wavelengths, we calculated the colour for three angular resolutions through the major axis of the disc, i.e. $2$ ~arcsec (a total of $110$ positions), $4$~arcsec (a total of $55$ positions), and $17$~arcsec (a total of $13$ positions). The [$3.6$]$-$[$4.5$] colours obtained through the disc is shown in Figure~\ref{fig:colres}. 

We calculated the stellar mass, $M_{\star}$~[$M_{\odot}$] at each position using the F$_{3.6\micron}$ and F$_{4.5\micron}$ fluxes (see Section~\ref{sec:farnear}) and the expression below \citep{eskew12};

\begin{equation}
 M_{\star}= 10^{5.65}\,\times\,(F_{3.6\micron})^{2.85}\times\,(F_{4.5\micron})^{-1.85}\times\,\frac{D^{2}}{{\rm 0.05^{2}}},
\label{eq:stelmass}
\end{equation} 			

where $D$ is the distance (Mpc). We estimated the $M_{\star}$ also using the expression given in \citet{que15} and obtained very similar $M_{\star}$ values to the ones estimated using the Eq.~\ref{eq:stelmass} above.

\section{Results and discussion}
\label{sec:resdis}

\subsection{Moment Maps and PVDs}
\label{pvdss}

The distribution of CO gas across the disc of NGC 1055 is shown in Figure~\ref{fig:pos}. As is usually the case, the $^{12}$CO(1--0) is more extended than its isotopologue $^{13}$CO(1--0) in NGC~1055, i.e. more than $50$~arcsec on each side of the disc (see Figure~\ref{fig:pos}). The PVDs for both lines are shown in the lower panels of Figure~\ref{fig:pos}. The PVD of $^{12}$CO(1--0) indicates that the gas reaches the maximum rotation velocity of the galaxy at about $50$ arcsec (equivalent to $3.6$~kpc) from the centre, and it is $213$~km~s$^{-1}$ (inclination correction applied), which is close to the value estimated from H\,{\small I} (HyperLeda; $\approx~182$~km~s$^{-1}$). 

An X-shaped PVD is usually seen in barred galaxies when they are edge-on, i.e. the central rapidly rising and outer slowly rising velocity components \citep{at99,bu99}. Ever since NGC 1055 is an edge-on barred galaxy, we would expect to see an X-shaped feature. However, there is only one velocity component visible in the $^{12}$CO(1--0) PVD of NGC~1055 (see Fig.~\ref{fig:pos}). It is interesting not to see the X-shaped feature in NGC~1055 which is a nearly edge-on barred galaxy (i.e. $i$~=~62.7$^\circ$). One explanation could be that the gas is possibly less bright, i.e. below the detection threshold, outside the central velocity component beyond the co-rotation.

As illustrated by the $^{12}$CO(1--0) moment 0 map and the PVD in Figure~\ref{fig:pos}, the CO gas exhibits a more extended distribution on the eastern side of the disc, reaching its maximum rotational velocity at $\approx60$~arcsec from the centre. However, the CO gas reaches its maximum rotational velocity at $\approx40$~arcsec on the western side of the disc. This suggests that the CO gas is more elongated to the east than to the west in the galaxy's disc. 

\subsection{Intensities and mass}
\label{sec:intensi}

The integrated CO line intensities decrease as the radius increases (see Fig.~\ref{fig:massinten}). Interestingly, the median $^{12}$CO(1--0) integrated intensity for the positions located in the SE (i.e. $19.8\,\pm\,1.6$~K~km~s$^{-1}$) is higher than the NW (i.e. $14.0\,\pm\,1.5$~K~km~s$^{-1}$). The overall brightness of CO gas is higher in the SE of the galaxy than in the NW. $^{13}$CO(1--0) is also brighter and more extended in the SE. This is also evident from the fact that $^{13}$CO(1--0) integrated intensities at almost all positions in the NW (except positions~$7$ and $8$, see Fig.~\ref{fig:pos}) are upper limits, while $^{13}$CO(1--0) was detected through the SE (except only one position).

Using the CO integrated intensities, we calculated the beam-averaged $M_{{\rm H}_2}$ and $\Sigma_{\rm H_{2}}$ at all positions studied 
for an angular resolution of $17$~arcsec (see Section~\ref{sec:linemass} and Table~\ref{tab:litedat}). As seen from Figure~\ref{fig:massinten}, $M_{{\rm H}_2}$ and $\Sigma_{\rm H_{2}}$ decrease with the radius going outwards. However, there is also a drop in both parameters in the central position (i.e. the position 0, see Figures~\ref{fig:pos} and ~\ref{fig:massinten}) because of the assumed depression in $X_{\rm CO}$ at the centre (see Section~\ref{sec:linemass}). Some processes in the ISM could cause the depression in the centre compared to the rest of the disc, such as high excitation temperature, strong turbulence, density, level of ionization by cosmic rays, metallicity, and large velocity dispersion \citep{sta91, dow98, nara11, isra20}. The median $M_{{\rm H}_2}$ and $\Sigma_{\rm H_{2}}$ values for the positions located in the SE (i.e. $M_{{\rm H}_2}=(1.43\,\pm\,0.01)\times10^{8}$~M$_{\odot}$ and $\Sigma_{\rm H_{2}}= 86.3\,\pm\,7.2$~M$_{\odot}$~pc$^{-2}$) are higher than the medians found in the NW (i.e. $M_{{\rm H}_2}=(1.02\,\pm\,0.01)\times10^{8}$~M$_{\odot}$ and $\Sigma_{\rm H_{2}}= 61.2\,\pm\,6.4$~M$_{\odot}$~pc$^{-2}$). This indicates that the SE of the disc has higher $M_{{\rm H}_2}$ and $\Sigma_{\rm H_{2}}$.

\subsection{Molecular line ratios}
\label{sec:ratres}

As seen in Figure~\ref{fig:ratmap}, the ratio of the moment 0 map shows fluctuations within an aperture of about $100$~arcsec in diameter or equivalently $\approx7.2$~kpc, i.e. some locations over the disc have higher ratios (i.e. the bright spots in Fig.~\ref{fig:ratmap}). The reasons for the fluctuations could be; 1) lower S/N ratio of $^{13}$CO(1--0) data; 2) some selected positions possibly coincide with the inter-arm regions where the emission tends to be weaker; 3) star formation processes could be different at some locations.

\begin{figure}
\begin{center}
  \includegraphics[width=8.5cm,clip=]{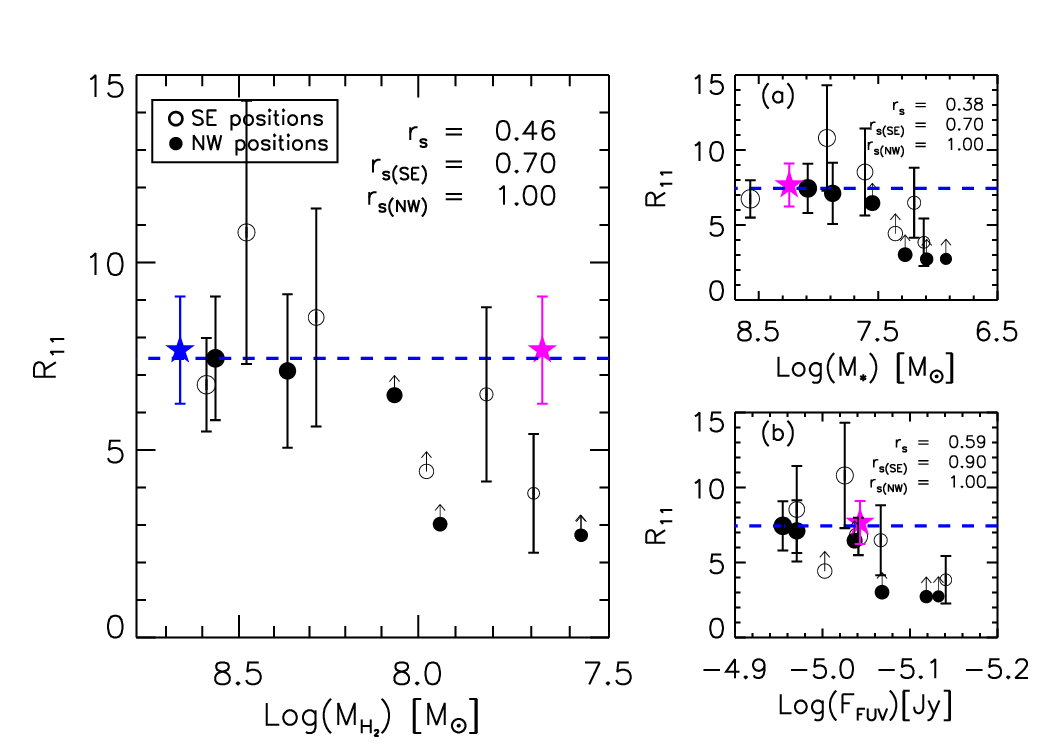}\\
  \includegraphics[width=8.5cm,clip=]{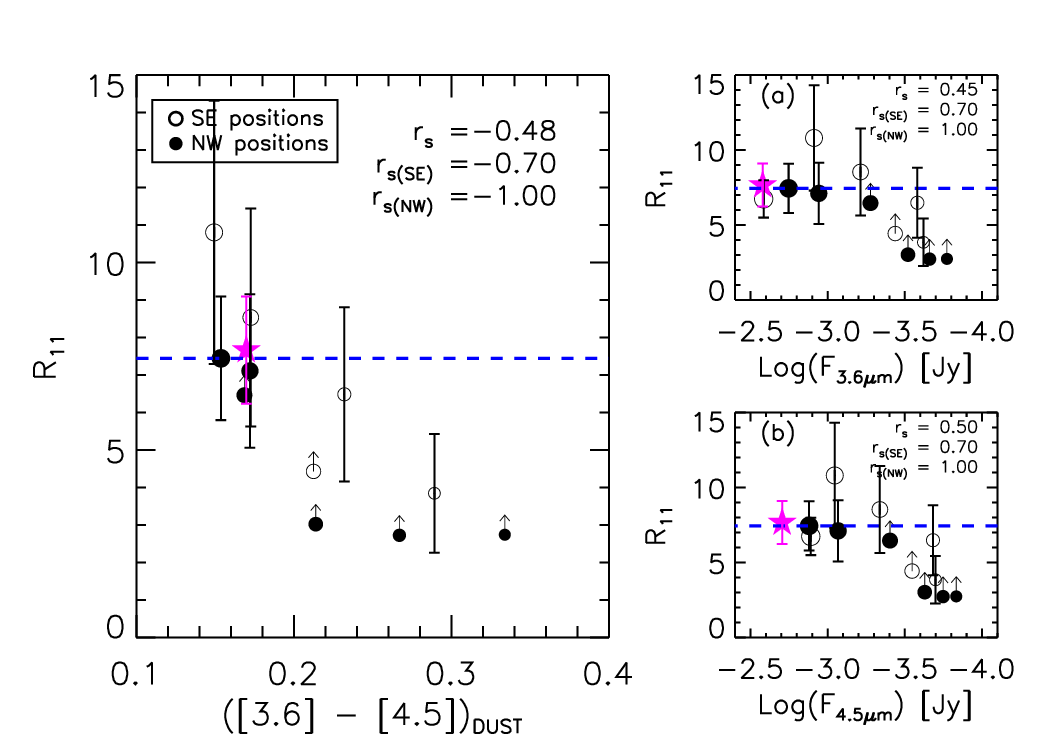}\\
  \caption{Top: The R$_{11}$ ratio versus $M_{{\rm H}_2}$ (left), $M_{\star}$ (panel $a$) and F$_{FUV}$ (panel $b$) is shown. Bottom: The R$_{11}$ ratio versus the [$3.6$]$-$[$4.5$] colour (left), F$_{3.6\micron}$ (panel $a$) and F$_{4.5\micron}$ (panel $b$) are shown. The median value of R$_{11}\,=\,7.44$ is shown by a dashed line in blue. The bigger the symbol size, the closer to the centre. The magenta stars show the value at the centre. The blue star symbol showing the highest $M_{{\rm H}_2}$ value in the top left panel represents the $M_{{\rm H}_2}$ at the central position in NGC~1055 as the magenta star, but it was calculated using the standard $X_{\rm CO}$ value for the Milky Way disc, rather than using the ten times lower $X_{\rm CO}$ (see Section~\ref{sec:linemass}). $r_{s}$, $r_{s\rm (SE)}$, and $r_{s\rm (NW)}$ represent the Spearman correlation coefficient for all positions, positions in the SE and NW, respectively, without considering lower limits values for the R$_{11}$ ratio (the data with arrows).}
    \label{fig:ratother}
\end{center}
\end{figure}

The R$_{11}$ ratios along the major axis of NGC~1055 have almost the same average and median value; i.e. $\approx7.3$ (see blue and red horizontal dashed lines in Figure~\ref{fig:lineratios}). The R$_{11}$ ratio ranges from $3.8$ to $10.8$ over the disc of NGC~1055 (excluding the lower limits, see Table~\ref{tab:ratios}), mostly lower than the values found for the arm and inter-arms of nearby spiral galaxies, i.e. R$_{11}\,=\,8.5 - 14.4$ \citep{cor18}. The R$_{11}$ ratios also show some fluctuations (i.e. local maxima or minima) over the disc of NGC~1055 (see Fig.~\ref{fig:lineratios}) with an overall decrease from the nucleus to the outskirts as seen in spiral galaxies, and also in gaseous lenticulars \citep[e.g.][]{pag01, sch10, tan11, topal16, topal20}.

The CO ratios in the central region of NGC~1055 have been analyzed in the past. \citet{isra20} found R$_{21}=\,0.71\,\pm\,0.20$ and \citet{mao10} found R$_{31}=\,0.40\,\pm\,0.03$ for the central $22$~arcsec. Our measurements for the central $17$~arcsec reveal R$_{11}=\,7.7\,\pm\,1.4$ (see Table~\ref{tab:ratios}). The central position of NGC~1055 has R$_{11}$ ratio at the lower end of the range found for the centre of the Milky Way and nearby spirals (R$_{11}\,=\,5$ - $27$), starbursts (R$_{11}\,=\,8.4$ - $15.7$), and Seyfert galaxies, i.e. R$_{11}\,=\,8.2$ - $16.2$ \citep{pag01, is09a,is09b,kri10}. The R$_{21}$ found in the central region of NGC~1055 is close to the median value of the range found in nearby galaxies; for spiral galaxies R$_{21}\,=\,0.6$ - $1.0$, for starbursts R$_{21}\,=\,0.6$ - $1.1$, and for Seyfert galaxies R$_{21}\,=\,0.7$ - $1.0$  \citep{ler09,is09a,is09b,koda20,debrok21,ler22}. Given that higher R$_{11}$ and R$_{21}$ ratios indicate thinner gas and higher gas kinetic temperature, respectively, the ratios found in the central position of NGC~1055 suggest that the galaxy hosts a central gas component with a similar gas kinetic temperature but possibly a higher gas density compared to the centre of spirals, starbursts, and Seyferts.

%
\begin{figure}
\begin{center}
  \includegraphics[width=8.0cm,clip=]{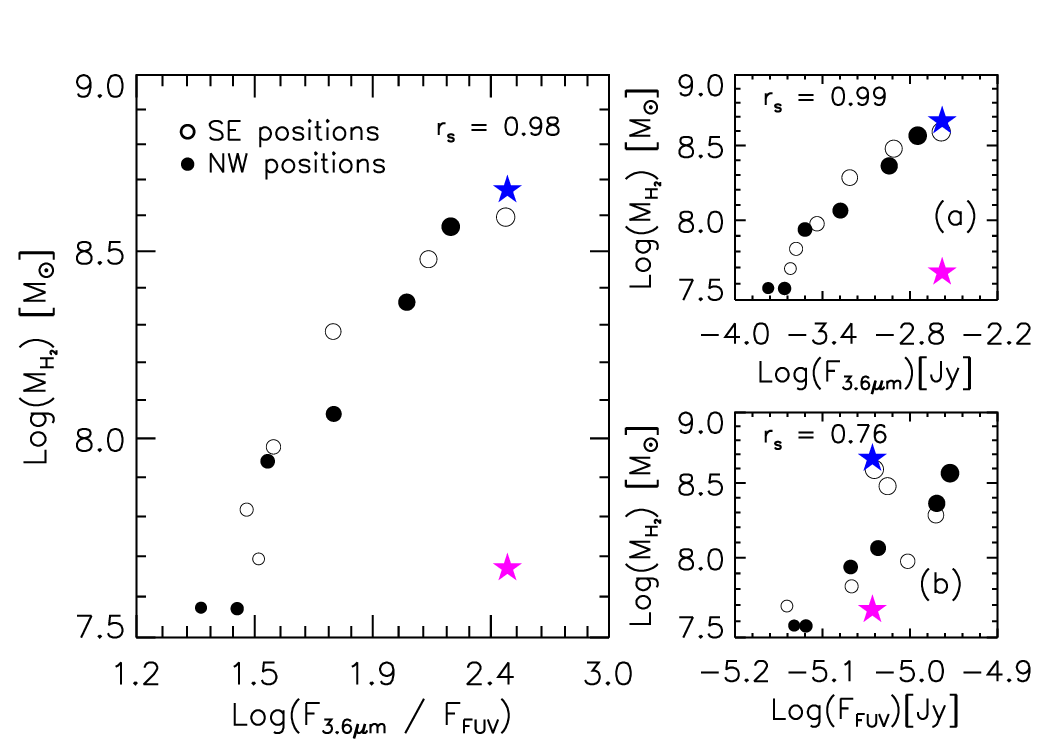} \\
   \includegraphics[width=8.0cm,clip=]{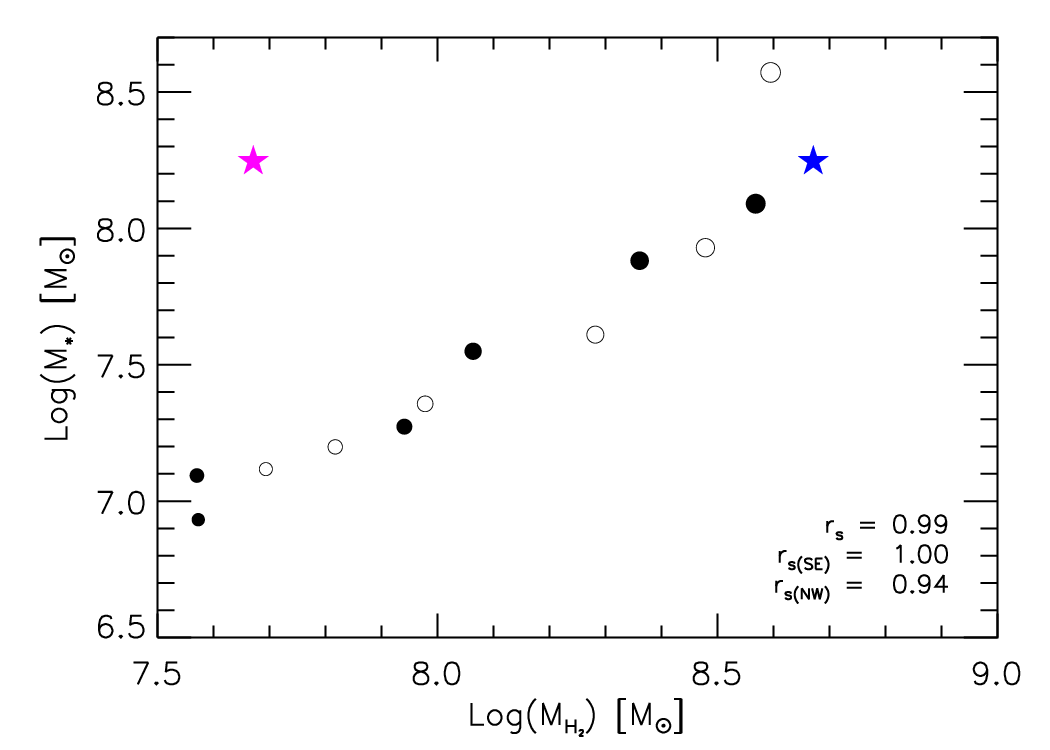} \\
  \caption{Top: The $M_{{\rm H}_2}$ is shown as a function of the F$_{FUV}$, F$_{3.6\micron}$, and F$_{3.6\micron}$/F$_{FUV}$ ratios (i.e. extinction) for an angular resolution of $17$~arcsec, i.e. the resolution of CO data. The small panels $a$ and $b$ show the $M_{{\rm H}_2}$ as a function of F$_{3.6\micron}$ and F$_{FUV}$, respectively. Bottom: The $M_{{\rm H}_2}$ is shown as a function of the $M_{\star}$ at selected $13$ positions for an angular resolution of $17$~arcsec. The symbol size is arranged so that the smallest symbol is the farthest from the centre (i.e. positions~$1$ and $12$, see Fig.~\ref{fig:pos}) while the largest symbol is the closest to the centre (i.e. positions~$6$ and $7$, see Fig.~\ref{fig:pos}). The magenta stars show the value at the centre. The blue stars (i.e. the star symbol with the highest $M_{{\rm H}_2}$ value in the panels) also represent the $M_{{\rm H}_2}$ at the central position in NGC~1055, assuming the standard $X_{\rm CO}$ value as explained in the caption to Figure~\ref{fig:ratother}. $r_{s}$ values in the panels represent the Spearman correlation coefficient considering all positions except the centre with a substantially low value of $M_{{\rm H}_2}$ (see Section~\ref{sec:linemass}).}
    \label{fig:massother}
\end{center}
\end{figure}

\subsection{Correlations among the physical parameters at three angular resolutions}
\label{sec:massfuv}
To investigate the effects of angular resolution on the results and correlations among gas, dust, and stellar populations at a matched resolution, we obtained the NIR data at three angular resolutions, namely $2$, $4$, and $17$~arcsec, and the FUV data at two angular resolutions, namely $4$ and $17$~arcsec (see Section~\ref{sec:farnear} and Table~\ref{tab:litedat}). This approach allows us to study the correlations among the three wavelength regimes, namely CO, FUV, and NIR, at a common beam size of $17$~arcsec, as well as between FUV and NIR at a common beam size of $4$~arcsec. Furthermore, it allows for investigating the associated physical properties of the three wavelength regimes at their original angular resolution, such as line ratios, molecular gas mass, stellar mass, extinction, and colour. 

The [$3.6$]$-$[$4.5$] colour obtained at three angular resolutions through the disc are shown in Figure~\ref{fig:colres}. The distribution of the colour across the disc of NGC~1055 shows a slight change as a function of the resolution (Fig.~\ref{fig:colres}). It is noteworthy that for the $17$~arcsec resolution, the colour at the central position (i.e. magenta star) has [$3.6$]$-$[$4.5$]$\,<\,0.2$ (i.e. lower than the minimum value for diffuse dust, see Figure~\ref{fig:colres}), while for resolutions of $4$ and $2$~arcsec [$3.6$]$-$[$4.5$]$\,>\,0.2$ at the central position. In addition, at the $17$~arcsec resolution, one position exhibits the colour [$3.6$]$-$[$4.5$] greater than $0.3$, while all other positions at all three resolutions consistently demonstrate the colour [$3.6$]$-$[$4.5$] less than $0.3$ (see Fig.~\ref{fig:colres} and Section~\ref{sec:reso2}).

For visualisation purposes, the y-axis in Figure~\ref{fig:colres} has been fixed. However, it is important to note that one data point associated with position~$6$ shows a negative colour of [$3.6$]$-$[$4.5$]$\,=-0.27$ for the $17$~arcsec resolution. This could indicate that the old stellar populations could be dominating the emission at $3.6\,\micron$ at position $6$, i.e. the increase in CO absorption possibly causes the $4.5\,\micron$ emission to be fainter than $3.6\,\micron$, opposite of what we see at all the other positions studied. Or this could simply be a result of the angular resolution, i.e. averaging over a large area. The following subsections discuss the correlations among the physical parameters for three angular resolutions. 

The Spearman correlation coefficients ($r_{s}$) were also calculated to assess the degree of correlation between the parameters. It should be noted that, for an angular resolution of $17$~arcsec, our initial approach did not include consideration of the central $M_{{\rm H}_2}$ value in any correlations involving $M_{{\rm H}_2}$ (i.e. top left panel in Figure~\ref{fig:ratother} and all panels in Figure~\ref{fig:massother}). This was due to the significant decrease in $M_{{\rm H}_2}$ at the centre resulting from the assumed depression in $X_{\rm CO}$ in the centre of the galaxy (see Section~\ref{sec:linemass}). However, to test the possible effects of the central depression in the gas mass on the correlations, we also included the central $M_{{\rm H}_2}$ value in the calculation of the Spearman correlation coefficients. The inclusion of the central $M_{{\rm H}_2}$ value did not affect the results considerably. In most cases, the correlations were degraded from very strong (i.e. $0.80$\,$<$\,$r_{s}$\,$<$\,$1.0$) to strong (i.e. $0.60$\,$<$\,$r_{s}$\,$<$\,$0.79$). The correlation was changed from weak (i.e. $r_{s}$\,$=$\,$0.46$) to very weak (i.e. $r_{s}$\,$=$\,$0.17$) for the R$_{11}$ ratio and $M_{{\rm H}_2}$ (top left panel in Figure~\ref{fig:ratother}), while the correlation changed the least for $M_{{\rm H}_2}$ and F$_{FUV}$, i.e. from $r_{s}$\,$=$\,$0.76$ to $r_{s}$\,$=$\,$0.70$ (panel b in Figure~\ref{fig:massother}).

Additionally, the $M_{{\rm H}_2}$ at the centre of NGC~1055 (also depicted in the associated figures) was calculated using the same $X_{\rm CO}$ value considered for the disc of the galaxy to ascertain the impact on the correlations. The results indicate that the correlations discussed below remain unaltered when utilising the standard Milky Way value for $X_{\rm CO}$ at the central position in NGC~1055. Overall, it should be noted that whether the central $X_{\rm CO}$ value is taken to be the standard Milky Way value (i.e. $2\times10^{20}$) or 10 times lower, the correlations involving $M_{{\rm H}_2}$ remain largely unaffected, with or without the central $M_{{\rm H}_2}$ value.

\subsubsection{Resolution of $17$~arcsec: The properties of molecular clouds in comparison to the other parameters}
\label{sec:reso17}
To compare the molecular gas-related properties, such as CO ratios and $M_{{\rm H}_2}$, to other physical properties, such as $M_{\star}$, F$_{FUV}$, F$_{3.6\micron}$, F$_{4.5\micron}$, and [$3.6$]$-$[$4.5$] colour, the FUV and NIR data were convolved to a common beam size of $17$~arcsec (see Sections~\ref{sec:intenmass} - \ref{sec:colmass}, and Table~\ref{tab:litedat}). 

The R$_{11}$ ratio as a function of $M_{{\rm H}_2}$, $M_{\star}$ and F$_{FUV}$ is shown in top panels of Figure~\ref{fig:ratother}. Considering all positions, there is weak (i.e. R$_{11}$ versus $M_{{\rm H}_2}$ and $M_{\star}$) and moderate correlations (i.e. R$_{11}$ versus F$_{FUV}$) between the pairs of parameters (i.e. $r_{s}\,=\,0.38 - 0.59$). However, the correlation is stronger in the SE, where the strongest correlation is between the R$_{11}$ and F$_{FUV}$ (i.e. $r_{s}\,=\,0.90$, see panel $b$ in the top panels of Fig.~\ref{fig:ratother}). Note that since the R$_{11}$ ratio at four positions in the NW is a lower limit, they were not considered for the calculation of the $r_{s}$. It is, therefore, natural to find a strong correlation in the NW, i.e. $r_{s}\,=\,1.00$, as it is based on the data at only two positions. 

All in all, the general trend indicates that all three parameters, i.e. $M_{{\rm H}_2}$, $M_{\star}$ and F$_{FUV}$, show a weak-to-moderate positive correlation with the R$_{11}$ ratio; they decrease in parallel with the ratio from the centre to the outskirts (the bigger the symbol size, the closer to the centre of the galaxy, see Fig.~\ref{fig:ratother}). Since the gas gets thicker (i.e. the R$_{11}$ decreases) from the centre to the outskirts, the number density of the young massive stars, as well as the mechanical feedback in the medium could be decreasing, supported by the decrease in the F$_{FUV}$ radially outwards. However, this general decrease of the R$_{11}$ with $M_{{\rm H}_2}$, $M_{\star}$ and F$_{FUV}$ holds for the SE but it is rather uncertain for the NW positions.

The R$_{11}$ ratio as a function of [$3.6$]$-$[$4.5$] colour, F$_{3.6\micron}$ (panel $a$) and F$_{4.5\micron}$ (panel $b$) is shown in the lower panels of Figure~\ref{fig:ratother}. The R$_{11}$ ratio and colour show a negative correlation (Fig.~\ref{fig:ratother}), while the correlations are positive between the R$_{11}$ ratio and NIR emissions. This indicates that the colour increases (becomes redder) while the gas gets thicker. F$_{3.6\micron}$ and F$_{4.5\micron}$ fluxes decrease (the dust gets fainter) with the R$_{11}$ ratios (the lower panels $a$ and $b$ in Fig.~\ref{fig:ratother}). The strongest correlation exists in the SE of the disc (neglecting the strong correlation found in the NW based on two data points only). Overall, for $17$~arcsec resolution, there is a decrease in dust emission, the R$_{11}$ ratios, and FUV emission, and the colour becomes redder when going radially outwards (Figure~\ref{fig:ratother}). 

Figure~\ref{fig:massother} shows that the $M_{{\rm H}_2}$ and F$_{3.6\micron}$/F$_{FUV}$ ratio are strongly correlated ($r_{s}=0.98$). A similar positive correlation exists between $M_{{\rm H}_2}$ and F$_{3.6\micron}$ (i.e. $r_{s}=0.99$) and a slightly weaker correlation between $M_{{\rm H}_2}$ and F$_{FUV}$ (i.e. $r_{s}=0.76$). Since the F$_{FUV}$ data have larger fluctuations (particularly for the positions in the SE) than the F$_{3.6\micron}$ data, this could cause the weaker correlation seen between $M_{{\rm H}_2}$ and F$_{FUV}$ compared to that between $M_{{\rm H}_2}$ and F$_{3.6\micron}$. As seen in the Figure, there is a robust positive correlation between $M_{\star}$ and $M_{{\rm H}_2}$ (i.e. $r_{s}\,=\,0.99$, excluding the centre). Both values exhibit a decline from the centre to the outskirts of the galaxy.

\subsubsection{Resolution of $4$~arcsec: The extinction, dust and FUV radiation}
\label{sec:reso4}

\begin{figure}
\begin{center}
   \includegraphics[width=8.5cm,clip=]{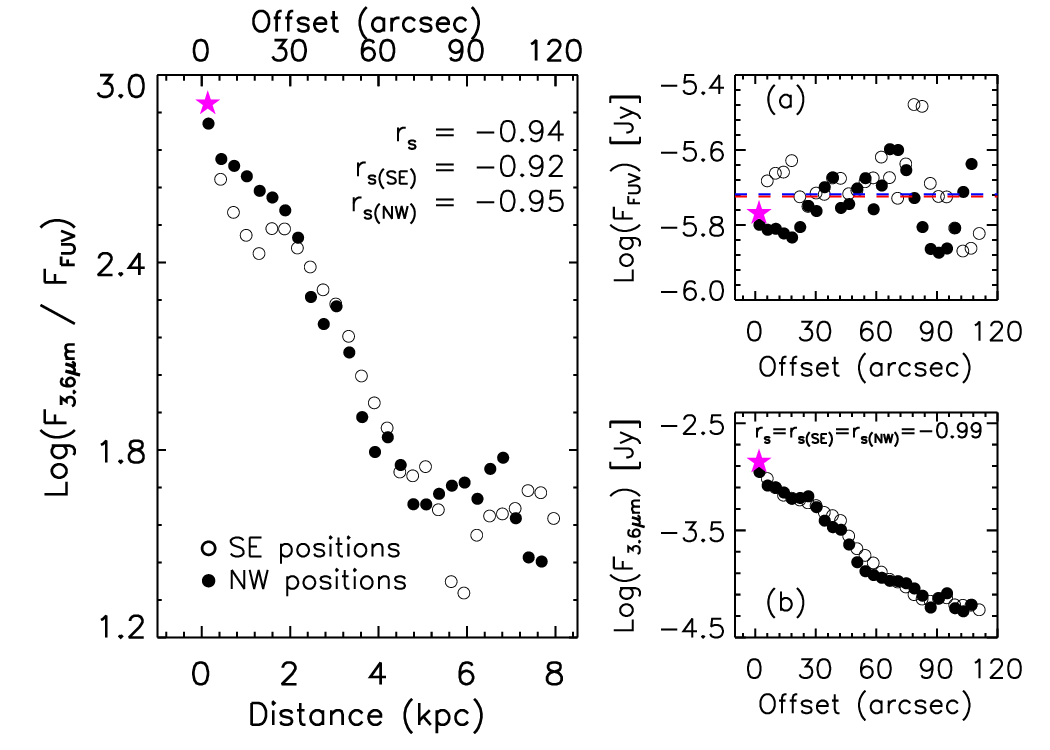}\\
  \caption{The radial variation of the F$_{3.6\micron}$/F$_{FUV}$ ratios, the F$_{FUV}$ and F$_{3.6\micron}$ fluxes obtained at $4$~arcsec resolution are shown. The red and blue dashed lines in panel $a$ are for the average and median values, respectively. In all panels, black circles represent the values in the NW (filled) and SE (open), while the magenta stars show the value at the centre. The correlation coefficients, i.e. $r_{s}$, are also shown in the panels.}
    \label{fig:fuvnir}
\end{center}
\end{figure}

%
%
%
\begin{figure}
\centering
      \includegraphics[width=8cm,clip=]{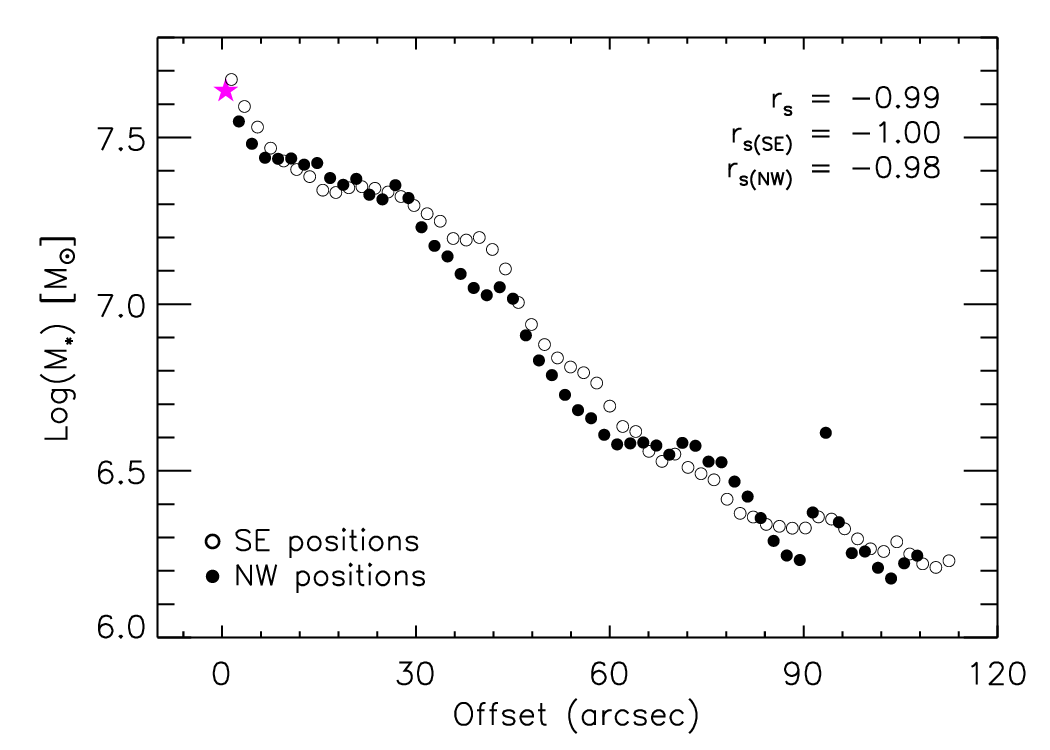} \\
      \includegraphics[width=8cm,clip=]{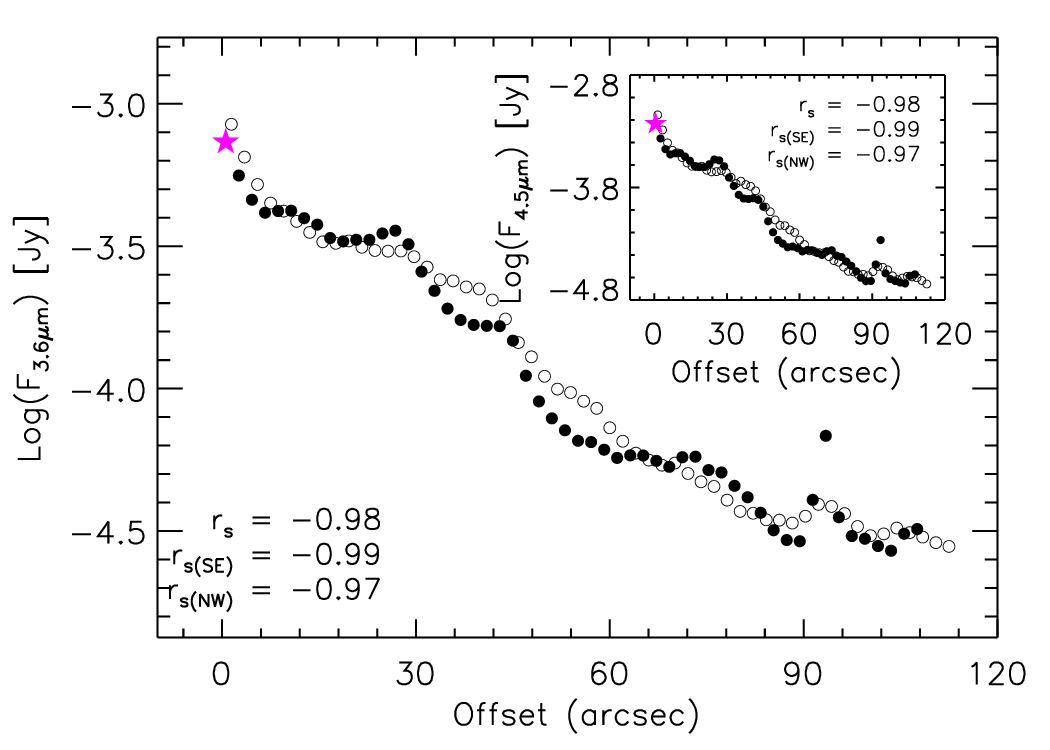}\\
  \caption{The $M_{\star}$ (top) and F$_{3.6\micron}$ along with F$_{4.5\micron}$ flux densities (bottom) calculated for an angular resolution of $2$ arcsec ($144$~pc) are shown. In all panels, filled and open black circles represent the values in the NW and SE, respectively, while the magenta stars show the value at the centre. The values for the Spearman correlation coefficients, i.e. $r_{s}$,  $r_{s(\rm SE)}$, and  $r_{s(\rm NW)}$ are shown on all panels.}
    \label{fig:steinfra}
\end{figure}

Figure~\ref{fig:fuvnir} shows the F$_{3.6\micron}$/F$_{FUV}$ ratio (i.e. extinction), F$_{FUV}$ and F$_{3.6\micron}$ fluxes as a function of galactocentric distance at a resolution of $4$~arcsec. The correlation is negative and strong between the extinction and galactocentric distance (i.e. $r_{s}=-0.94$), i.e. the extinction decreases as the distance increases, reaching the lowest value at $\approx6$~kpc in the SE (left panel in Figure~\ref{fig:fuvnir}). However, in addition to this general trend, three prominent areas across the disc separate the eastern and western sides by showing differences in the variation of the extinction and F$_{FUV}$. Firstly, from the centre to an offset of $\approx\,20$~arcsec (equivalently $1.4$~kpc), the decrease in the extinction is slightly steeper in the SE compared to the NW. Within the same central area, while the F$_{FUV}$ steeply increases in the SE, it decreases in the NW (panel~$a$ in Fig.~\ref{fig:fuvnir}). 

Secondly, from $20$ to $70$~arcsec the extinction decreases (the left panel in Fig.~\ref{fig:fuvnir}) while the F$_{FUV}$ shows fluctuations around the median on both sides of the disc (panel~$a$ in Fig.~\ref{fig:fuvnir}). However, almost all values of F$_{FUV}$ in the SE are higher than the median compared to the values in the NW (panel~$a$ in Fig.~\ref{fig:fuvnir}). 

Finally, after $\approx\,70$~arcsec (equivalently $5.0$~kpc), the extinction suddenly increases between $70$ and $90$~arcsec followed by a decrease in the NW. However, the extinction reaches its minimum in the SE at $\approx\,90$~arcsec, followed by a steep increase (the left panel in Fig.~\ref{fig:fuvnir}). The F$_{FUV}$ shows a sudden decrease followed by an increase after $70$~arcsec in the NW, while it reaches its maximum value in the SE at an offset of $80$~arcsec ($5.8$~kpc) and suddenly decreases (panel~$a$ of Fig.~\ref{fig:fuvnir}). The variation of the extinction and F$_{FUV}$ follows an opposite pattern after about $70$~arcsec in each side of the disc.

The change in the F$_{FUV}$ and extinction within the central zone of about $40$~arcsec in diameter ($\approx3$~kpc) and in the outskirts shows different trends on each side of the disc. The eastern part of the disc is mostly brighter in the FUV compared to the western part (Fig.~\ref{fig:fuvnir}). However, the F$_{3.6\micron}$ decreases with radius as indicated by a strong negative correlation (panel b Figure~\ref{fig:fuvnir}). The variation in the extinction as a function of the radius explained above is, therefore, mostly driven by the change in the F$_{FUV}$ showing notable fluctuation compared to the F$_{3.6\micron}$ across the disc (see small panels in Fig.~\ref{fig:fuvnir}). This indicates that the dust could be more uniformly distributed across the galaxy's disc than young stars located mostly in the arms causing fluctuations in the FUV data.

Overall, there is an inhomogeneity and asymmetry in the distribution of young massive stars across the disc. NGC~1055 is not a member of a pair or group, and there is no information in the literature about recent or past minor/major mergers or interactions with other galaxies. However, it is still likely that multiple accretions of gas from intergalactic space onto the disk could have occurred during the galaxy's lifetime. This scenario could explain the inhomogeneity and asymmetry seen in the stellar population across the disc.

%
\subsubsection{Resolution of $2$~arcsec: The colour, stellar mass and metallicity gradients}
\label{sec:reso2}

The radial change of $M_{\star}$, F$_{3.6\micron}$ and F$_{4.5\micron}$ for the resolution of $2$~arcsec are shown  in Figure~\ref{fig:steinfra}. There is a strong negative correlation between each parameter and the distance from the galactic centre (i.e. $r_{s}\,\approx\,-1$), with some minor fluctuations (Figure~\ref{fig:steinfra}). The correlation is equally strong for positions on either side of the disc, suggesting no asymmetry in this respect. 

ISRF is filled with many sources including synchrotron radiation, cosmic microwave background, and emission from dust, stars, and hot plasma. Since all positions have [$3.6$]$-$[$4.5$]$\,> 0$, the positions could be dominated by dust emission (heated by the ISRF, with a possible contribution from young massive stars). 

\begin{figure*}
\centering
      \includegraphics[width=15cm,clip=]{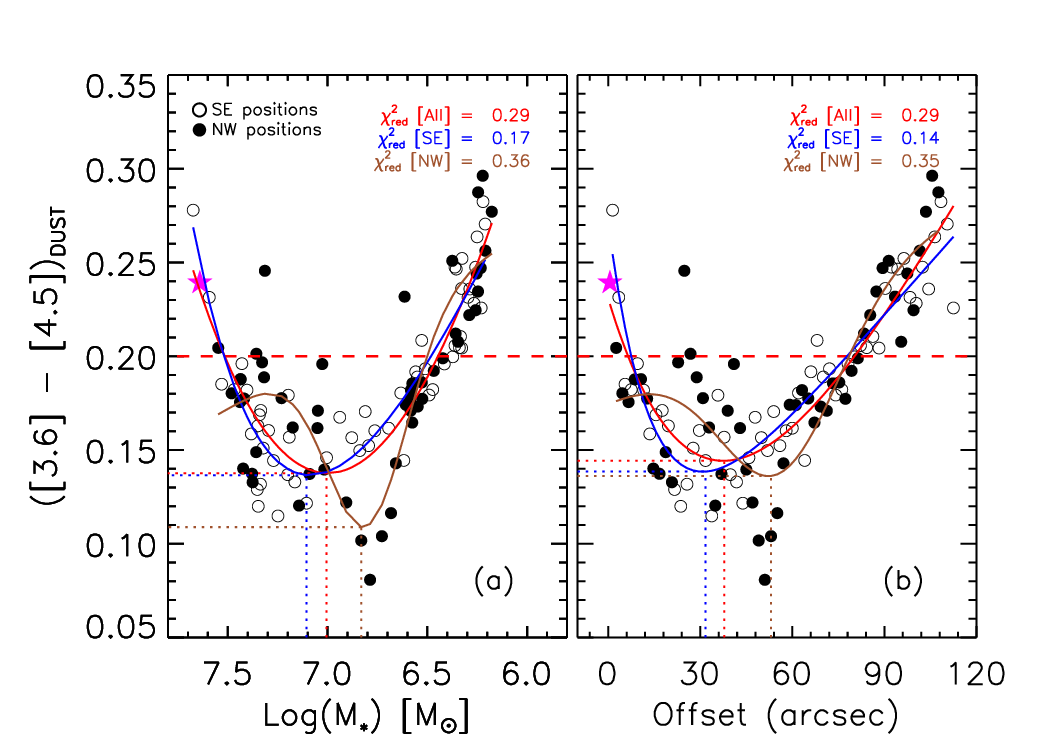}\\
  \caption{The [$3.6$]$-$[$4.5$] colour as a function of the $M_{\star}$ (panel a), and the angular distance from the galactic centre (panel b) are shown for $2$~arcsec ($144$~pc) resolution, i.e. the original angular resolution for $3.6\,\micron$ and $4.5\,\micron$ data. The magenta stars in each panel show the value at the centre. The horizontal dashed red line in the panels represents the lowest value for the colour of diffuse dust. The reduced chi-square ($\chi^2_{\rm red}$) values for three different fits (see Section~\ref{sec:reso2}) are represented with three different colours; for all positions (red), for the positions in the SE (blue) and NW (brown).}
    \label{fig:stecolmas}
\end{figure*}

The [$3.6$]$-$[$4.5$] colour as a function of $M_{\star}$ (panel $a$) and the galactocentric distance (panel $b$) are shown in Figure~ \ref{fig:stecolmas}. To estimate the minimum value for the [$3.6$]$-$[$4.5$] colour and corresponding $M_{\star}$ for the distribution, we performed a non-linear least-squares fit to the data using IDL procedure $gaussfit$. The fit consists of a combination of the Gaussian and linear regression functions. As shown in Figure~\ref{fig:stecolmas}, we performed the fit for three data sets separately, i.e. for all positions (i.e. the red line), for the positions in the SE (the blue line), and finally, for the positions in the NW (i.e. the brown line), to identify any differences along and each side of the disc. 

Considering all positions studied, our fit to the colour - $M_{\star}$ parameter space (red solid line in panel a of Fig.~\ref{fig:stecolmas}) indicates a minimum value of $0.138$ for the [$3.6$]$-$[$4.5$] colour and a corresponding value of log($M_{\star}$)\,=\,$7.01$ (red dotted lines in the same panel). We obtained a similar minimum value for the colour in the SE of the disc (blue solid line in panel a of Fig.~\ref{fig:stecolmas}), i.e. [$3.6$]$-$[$4.5$]\,=\,$0.136$, and the corresponding value of log($M_{\star}$)\,=\,$7.12$ (blue dotted lines in the same panel). However, based on the fit, the NW has the smallest value for the colour, i.e. [$3.6$]$-$[$4.5$]\,=\,$0.11$, and for $M_{\star}$, i.e. log($M_{\star}$)\,=\,$6.83$ (see brown solid and dotted lines in panel a of Fig.~\ref{fig:stecolmas}). 
 
The lowest observed value for the colour (the bluest) among all positions (i.e. $110$ positions, see Section~\ref{sec:colmass}) is located in the NW, reaching a minimum value of [$3.6$]$-$[$4.5$]\,=\,$0.08$ at an offset of $\approx51$~arcsec (panel b in Fig.~\ref{fig:stecolmas}). On the other hand, the bluest observed colour in the SE is  [$3.6$]$-$[$4.5$]\,=\,$0.115$ and located at an offset of $\approx34$~arcsec (panel b in Fig.~\ref{fig:stecolmas}). The fit to the colour - distance parameter space indicates that the colour reaches its minimum value of [$3.6$]$-$[$4.5$]\,=\,$0.138$ at $31.7$~arcsec (a linear size of $\approx2.3$~kpc) in the SE (i.e. the fit in blue in panel b of Fig.~\ref{fig:stecolmas}). The colour reaches a very similar minimum value of [$3.6$]$-$[$4.5$]\,=\,$0.136$ at an offset of $53.1$~arcsec (or $\approx3.8$~kpc) in the NW (see the fit in brown in panel b of Fig.~\ref{fig:stecolmas}). Considering the fit to all positions studied (red solid line in panel b of Fig.~\ref{fig:stecolmas}), the colour reaches its minimum value of [$3.6$]$-$[$4.5$]\,=\,$0.144$ at an offset of $37.8$~arcsec (red dotted lines in the same panel). 

Overall, the colour does not reach its minimum at the same offset on each side of the disc, i.e. the minimum takes place further in the NW than SE (Fig.~\ref{fig:stecolmas}). The variation of the colour in the SE is similar (i.e. the fit in blue in Fig.~\ref{fig:stecolmas}) to the variation across the whole disc (i.e. the fit in red in Fig.~\ref{fig:stecolmas}). The NW part of the disc shows a distinct feature in this regard, i.e. more scattered and has the bluest colour (i.e. the fit in brown in Fig.~\ref{fig:stecolmas}).

The [$3.6$]$-$[$4.5$] colour range found in NGC~1055 indicates non-stellar origin, i.e. [$3.6$]$-$[$4.5$]\,$>$\,$0$. In addition, PAHs usually have colours [$3.6$]$-$[$4.5$]\,$\sim$\,$0.3$, the typical value for PAH-dominated dust emission \citep{fla06}, and the colour becomes bluer as the PAHs become weaker. Since the colours obtained in the disc of NGC~1055 are always less than $0.3$ (see Fig.~\ref{fig:stecolmas}), PAHs could not be dominant in the dusty disc of the galaxy. The PAHs could be the least dominant at about $30$~arcsec in the SE and $50$~arcsec in the NW of NGC~1055, where the colour is bluest, compared to the rest of the disc (Fig.~\ref{fig:stecolmas}).

From the centre to $\approx4$~arcsec ($0.3$~kpc) and after $\approx80$~arcsec ($5.8$~kpc) on both sides of the disc, the colour is within the range for diffuse dust, i.e. $0.2$\,$<$\,[$3.6$]$-$[$4.5$]\,$<$\,0.7 (Fig.~\ref{fig:stecolmas}). However, all positions located between $4$ and $80$~arcsec in the disc (except only three positions) have colours $0$\,$<$\,[$3.6$]$-$[$4.5$]\,$<$\,0.2; the colour reaches a minimum at about $40$~arcsec and then starts to increase until it reaches a value of $0.2$ at about $80$~arcsec from the centre (Fig.~\ref{fig:stecolmas}). Overall, the colour gets bluer from the centre up to $40$~arcsec and then it gets redder through the disc going outwards.

The change in the [$3.6$]$-$[$4.5$] colour across the disc of NGC~1055 suggests that a mixture of dust and PAH emission is possible in a central region of $0.3$~kpc radius and after $5.8$~kpc on both sides of the disc. On the other hand, relatively less diffuse dust is likely to be located between $0.3$~kpc and $5.8$~kpc on both sides of the disc. This suggests that the galaxy is redder in the centre and at the outskirts, and relatively bluer in between.

For all galaxies on average, the metallicity decreases with the radius, but the gradient seen in the metallicity is the steepest in normal spirals while it gets flatter from barred spirals to lenticular and elliptical galaxies \citep{hen99}. The galaxy mass and metallicity \citep{tre04} and also the metallicity in the halo of spirals, where red giant stars are dominant, and the galaxy’s luminosity \citep{mou05} show a positive correlation. Spiral arms could have a higher metallicity as a result of more efficient star formation, while the metallicity in the inter-arm regions is relatively lower \citep[e.g.][]{ho17, ho18}. \citet{pel12} studied the [$3.6$]$-$[$4.5$] colour gradient in a sample of ETGs and found that the colour gets redder from the centre to the outskirts. They conclude that the colour gradient and metallicity are correlated, i.e. the galaxies are more metal-poor (redder colour) through the outskirts. The colour gradient across the disc of NGC~1055 indicates an increase in metallicity up to halfway through the gaseous disc as the [$3.6$]$-$[$4.5$] colour gets bluer followed by a decrease in metallicity going outwards as the [$3.6$]$-$[$4.5$] colour gets redder.

\section{Conclusion}
\label{sec:conc}

In this work, we have studied archival CO observations together with FUV and NIR data to investigate the correlations between the properties of stellar populations, gas, and dust over the disc of the spiral galaxy NGC~1055 at multiple angular resolutions. Our conclusions are as follows.

\begin{itemize}

\item $^{12}$CO(1--0) gas extends further over the disc, i.e. about $150$~arcsec or $\approx11$~kpc on each side of the disc, and is more extended in the SE compared to the NW, in contrast to $^{13}$CO(1--0) which is centrally concentrated. Additionally, $^{12}$CO(1--0) is brighter in the SE than in the NW, and median values for gas mass and gas surface densities in the SE are also higher than in the NW.

\item We find that the R$_{11}$ ratio [$^{12}$CO(1--0) / $^{13}$CO(1--0)] decreases with the galactocentric radius. The R$_{11}$ ratio map shows elevated values that indicate the presence of optically thin gas, compared to the rest of the molecular disc. R$_{11}$, R$_{21}$, and R$_{31}$ ratios in the centre exhibit similar gas kinetic temperature but possibly less tenuous and denser compared to the centre of spirals, starbursts, and Seyferts. 

\item Our results show a weak-to-moderate linear correlation between the R$_{11}$ ratio and the other physical properties, such as $M_{{\rm H}_2}$, $M_{\star}$, FUV and NIR radiation, and the [$3.6$]$-$[$4.5$] colour. The data for $17$~arcsec resolution ($1.2$~kpc) indicate that the R$_{11}$ ratio decreases (the gas gets thicker) along with the number of young massive OB stars, and the dust emission gets fainter going outwards in the disc. 

\item The $4$~arcsec ($288$~pc) data indicate that the FUV emission shows considerable fluctuations across the disc of the galaxy whereas the dust is distributed more uniformly. This could be explained by the fact that young massive stars, i.e. the source for FUV radiation, are mostly located in the arms of spiral galaxies causing the observed fluctuations. All indicate an inhomogeneity and asymmetry in the distribution of young stars across and also on either side of the disc. This could be a result of recent or past gas accretion to the disc.

\item Our results indicate the existence of two distinct regions in the disc of the galaxy; (1) the diffuse dust in the centre and the outskirts, and (2) relatively less diffuse dust located in between. As the $2$~arcsec ($144$~pc) data indicate, the colour reaches a minimum (becomes the bluest) halfway through the disc and then increases (becomes redder) going outwards through the disc. This indicates a fluctuation in the stellar populations and metallicity across the gaseous disc, while the $M_{\star}$ and NIR emissions show a more gradual decrease from the centre to the outskirts.

\end{itemize}

%
\section*{Acknowledgements}
The author thanks the anonymous referee for constructive comments and suggestions. This work is based [in part] on archival data obtained with the Spitzer Space Telescope, which was operated by the Jet Propulsion Laboratory, California Institute of Technology under a contract with NASA. We acknowledge the usage of the HyperLEDA database (http://leda.univ-lyon1.fr). This research has used the NASA/IPAC Extragalactic Database (NED), operated by the Jet Propulsion Laboratory, California Institute of Technology, under contract with the National Aeronautics and Space Administration. 

\section*{Data availability}
The data underlying this article will be shared on a reasonable request to the corresponding author.

\bibliographystyle{mn2e}
\bibliography{reference}
%
%
\appendix
%
%
\section{Integrated CO profiles and best fits}
\label{sec:Ap1}

\begin{figure*}
  \hspace{-15pt}
  \includegraphics[width=5.8cm,clip=]{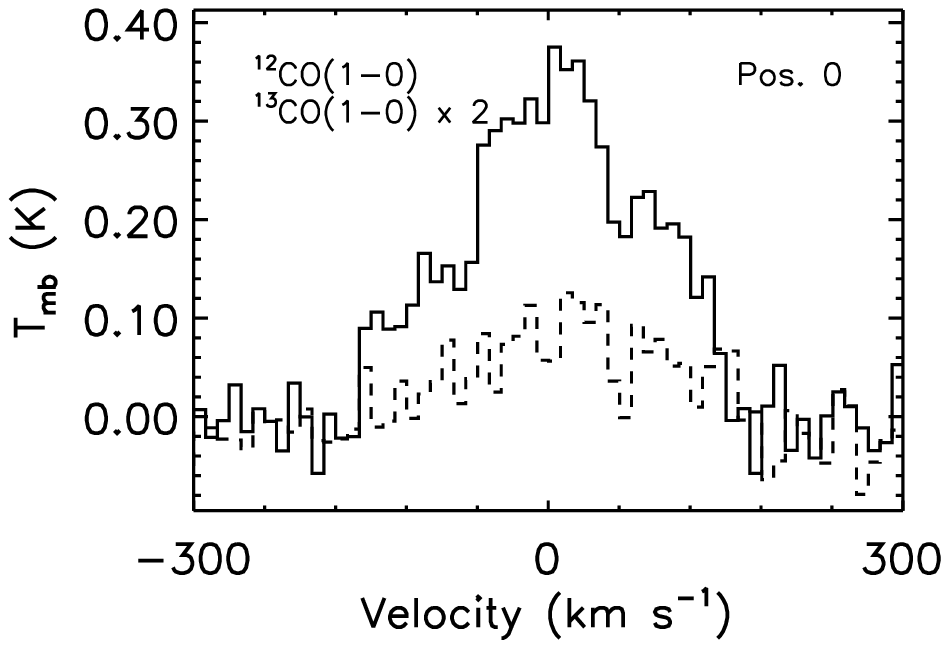}
  \includegraphics[width=5.8cm,clip=]{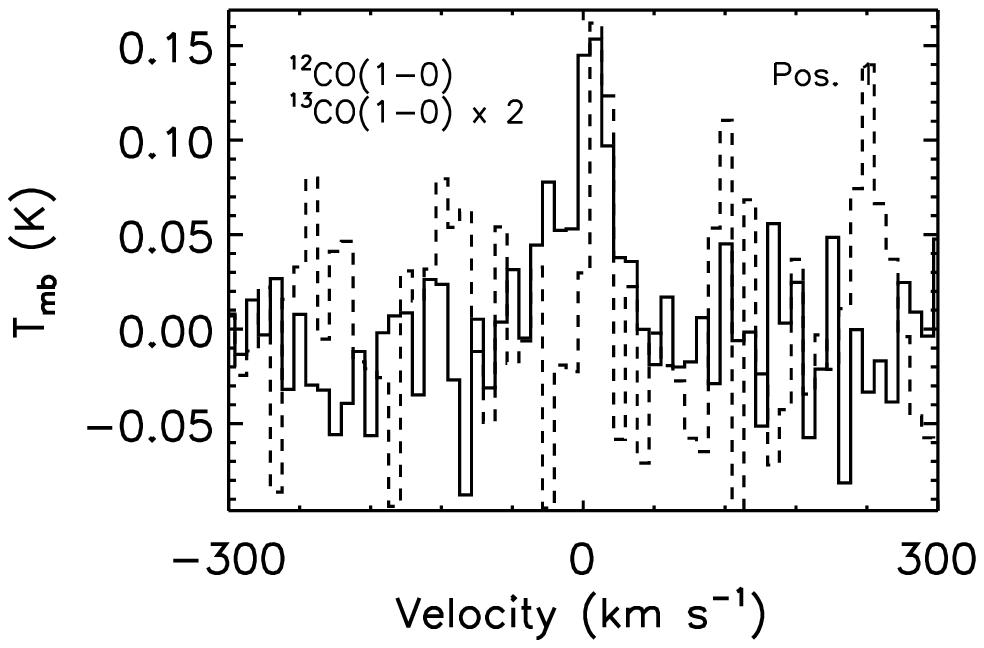}
    \includegraphics[width=5.8cm,clip=]{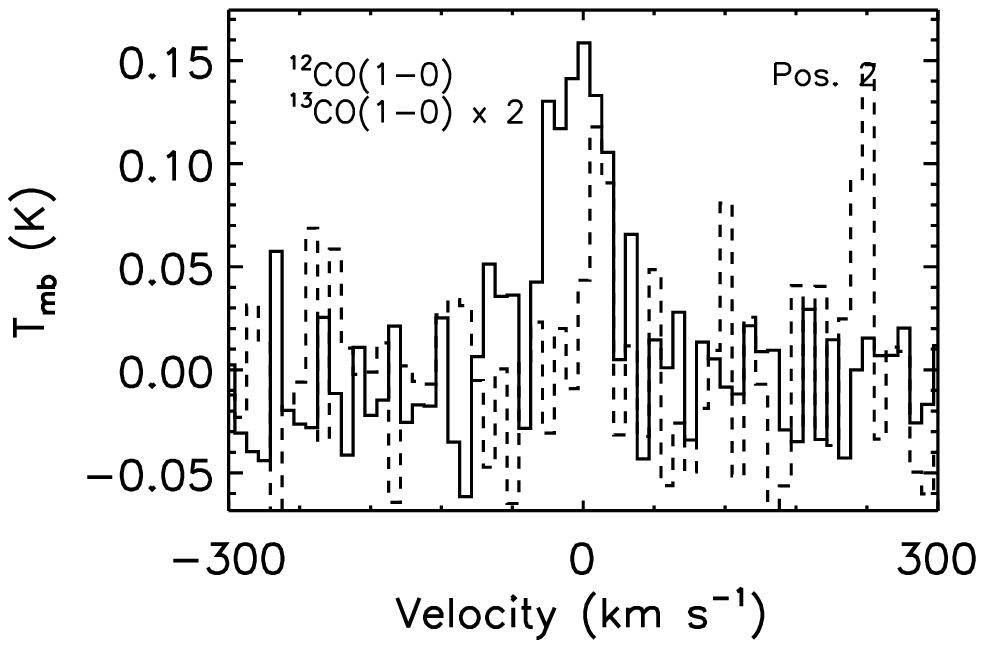}\\
  \vspace{-10pt}
  \hspace{-15pt}
  \includegraphics[width=5.8cm,clip=]{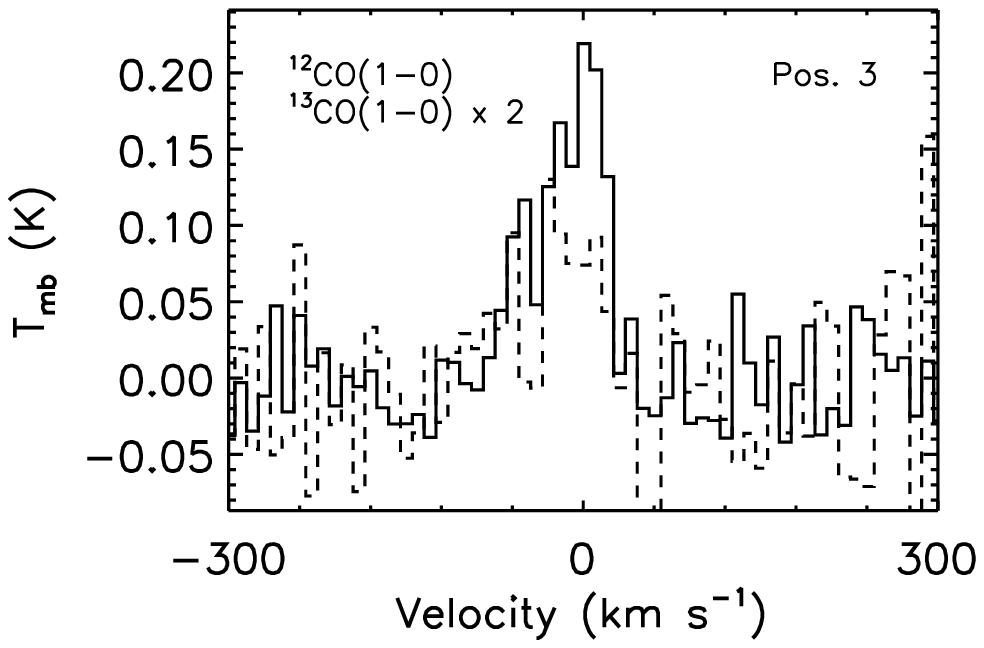}
  \includegraphics[width=5.8cm,clip=]{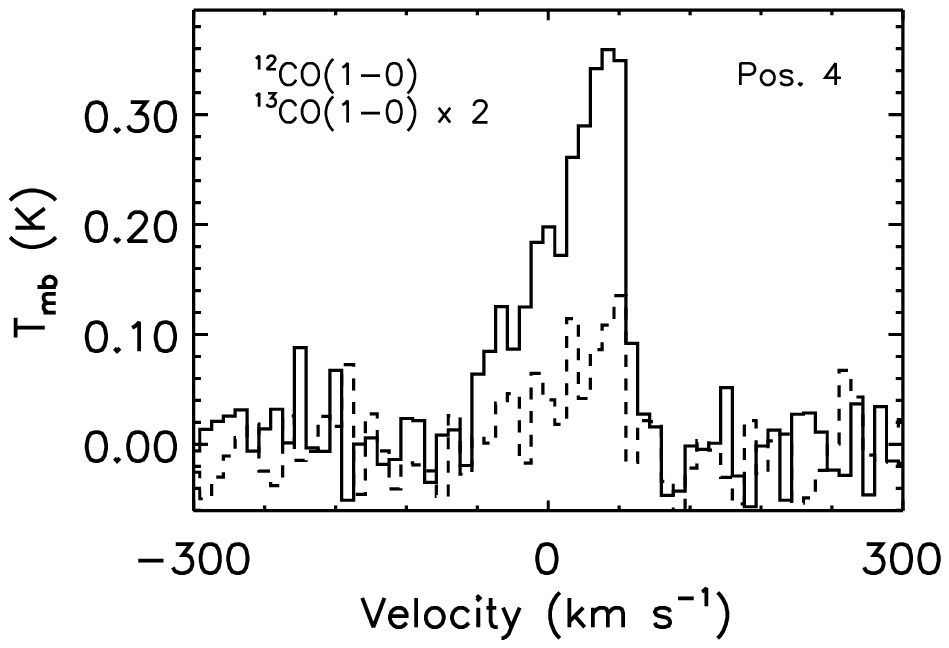}
    \includegraphics[width=5.8cm,clip=]{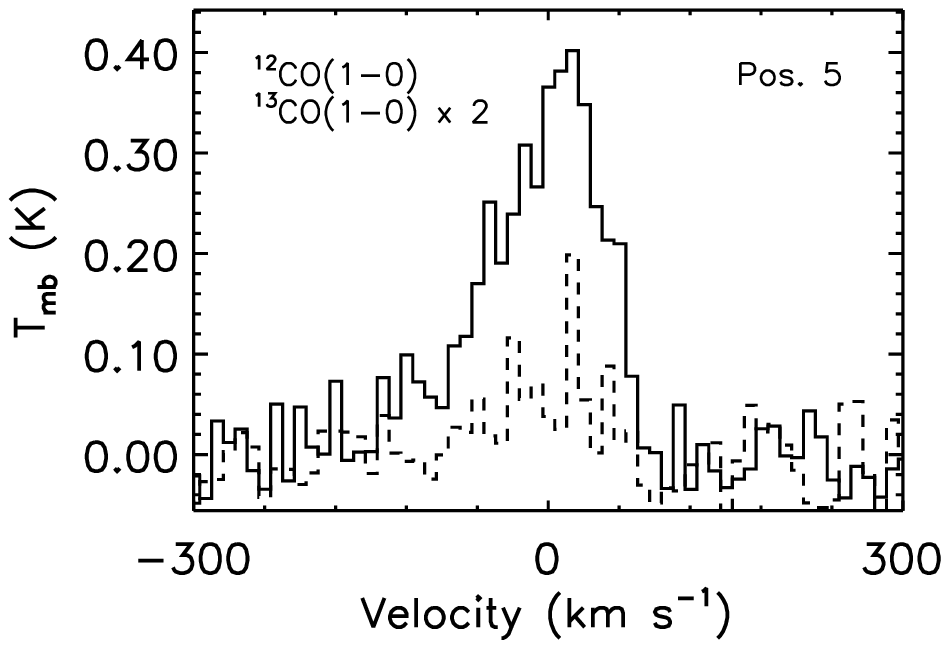}\\
  \vspace{-10pt}
  \hspace{-15pt}
  \includegraphics[width=5.8cm,clip=]{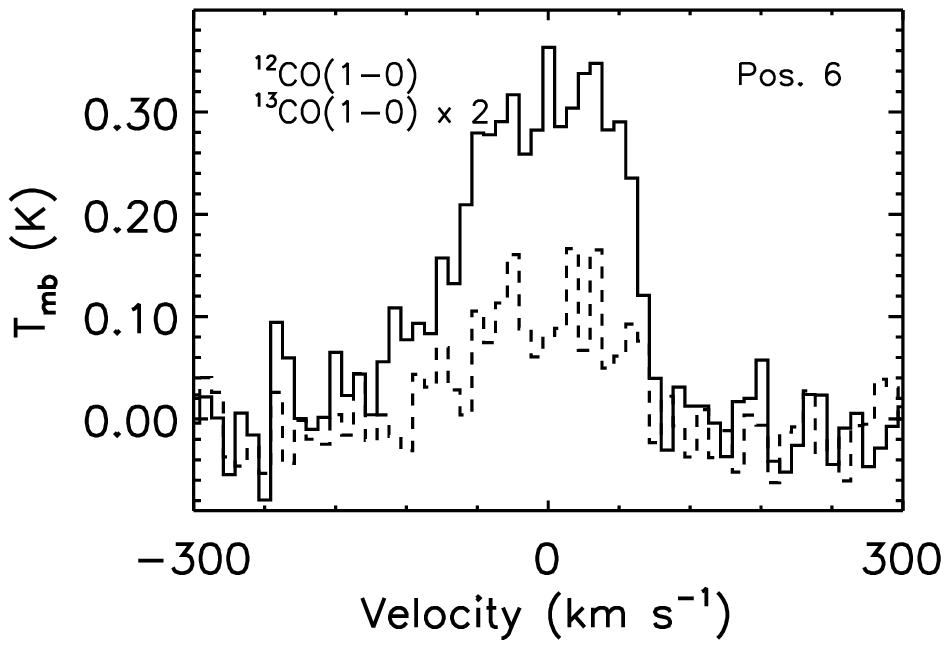}
  \includegraphics[width=5.8cm,clip=]{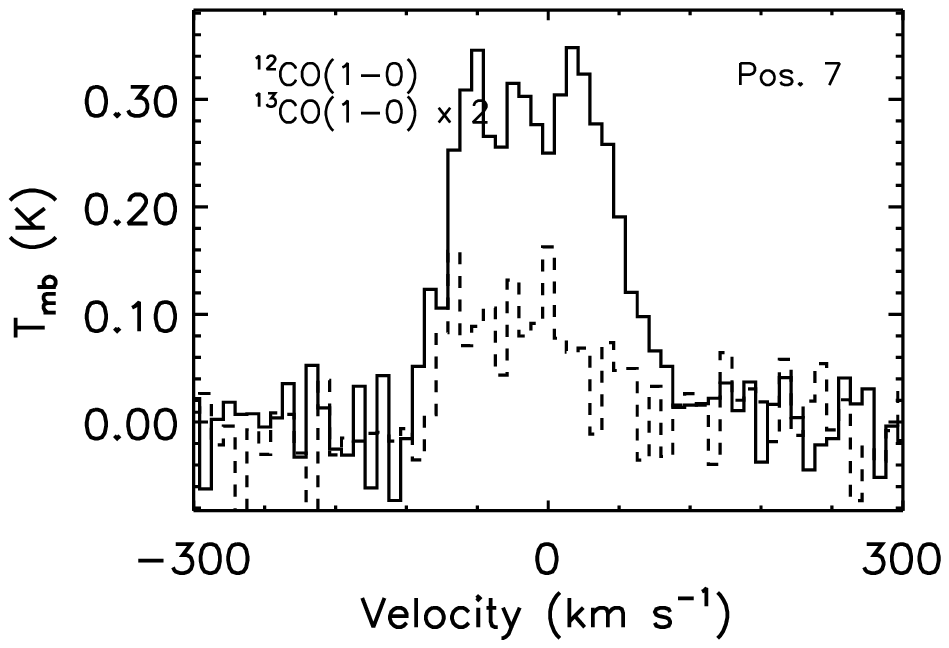}
    \includegraphics[width=5.8cm,clip=]{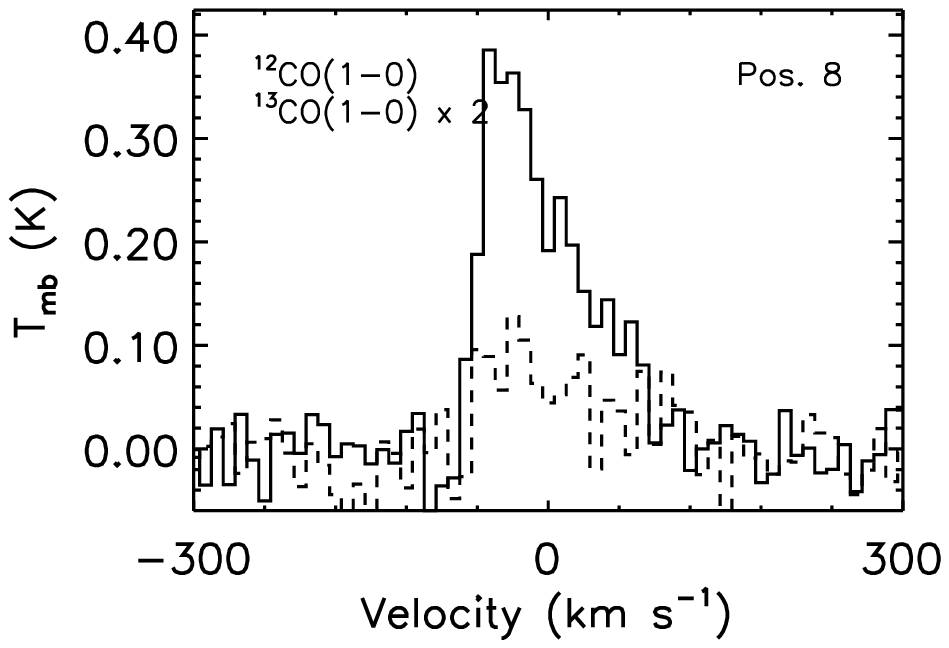}\\
      \vspace{-10pt}
  \hspace{-15pt}
  \includegraphics[width=5.8cm,clip=]{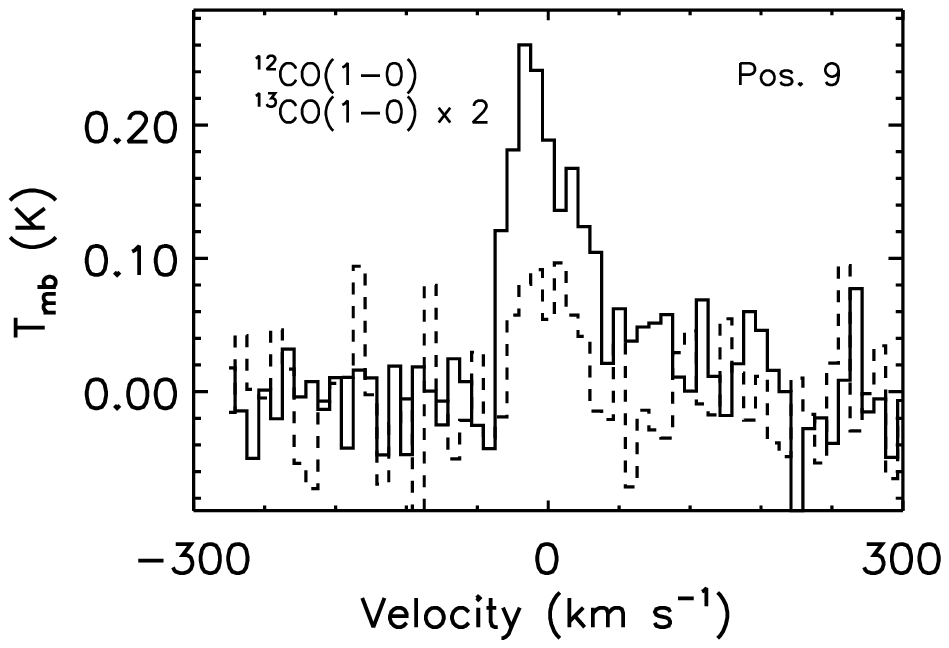}
  \includegraphics[width=5.8cm,clip=]{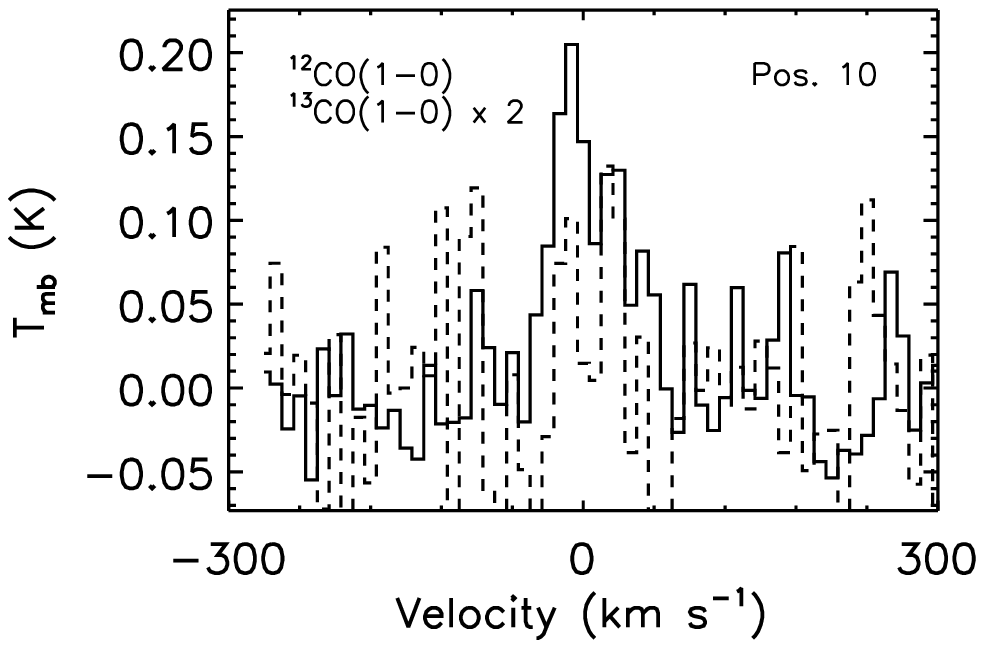}
    \includegraphics[width=5.8cm,clip=]{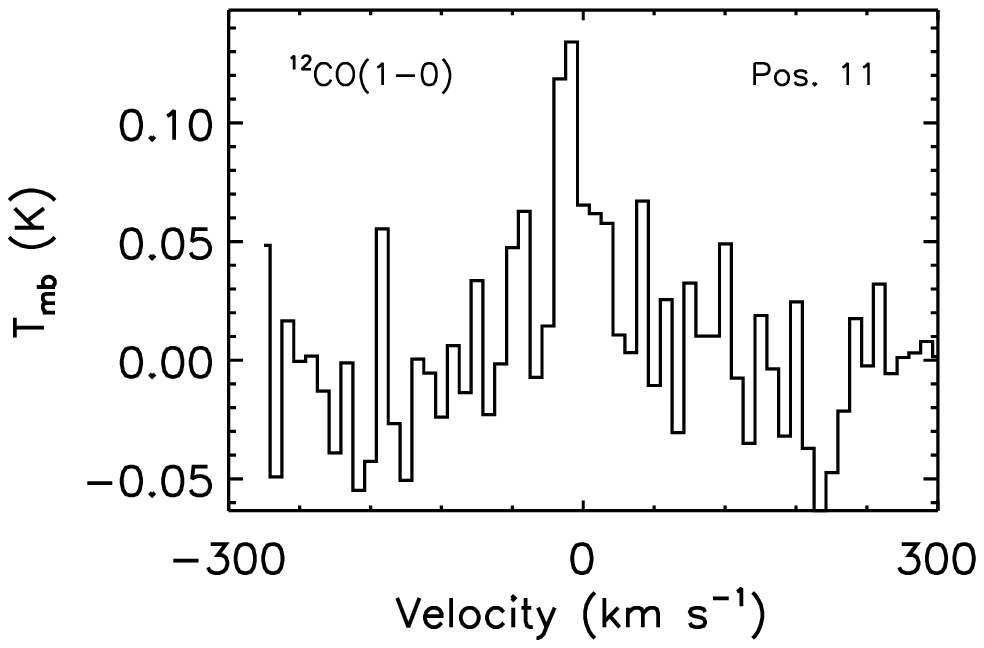}\\
      \vspace{-10pt}
  \hspace{-15pt}
    \includegraphics[width=5.8cm,clip=]{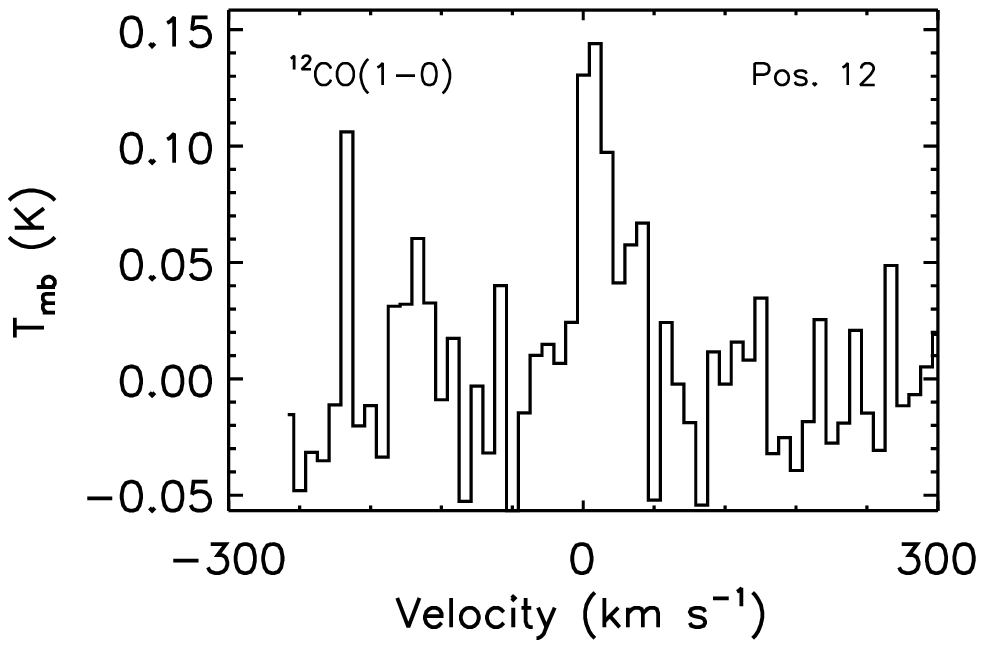}
      \caption{CO lines detected in the disc of NGC~1055. $^{13}$CO(1--0) line (dashed) is multiplied by $2$ to make them more visible. 
      The position numbers and names of emission lines are also indicated in each plot's top-right and top-left corners, respectively.}
  \label{fig:pro}
\end{figure*}


\begin{figure*}
  \hspace{-15pt}
  \includegraphics[width=5.8cm,clip=]{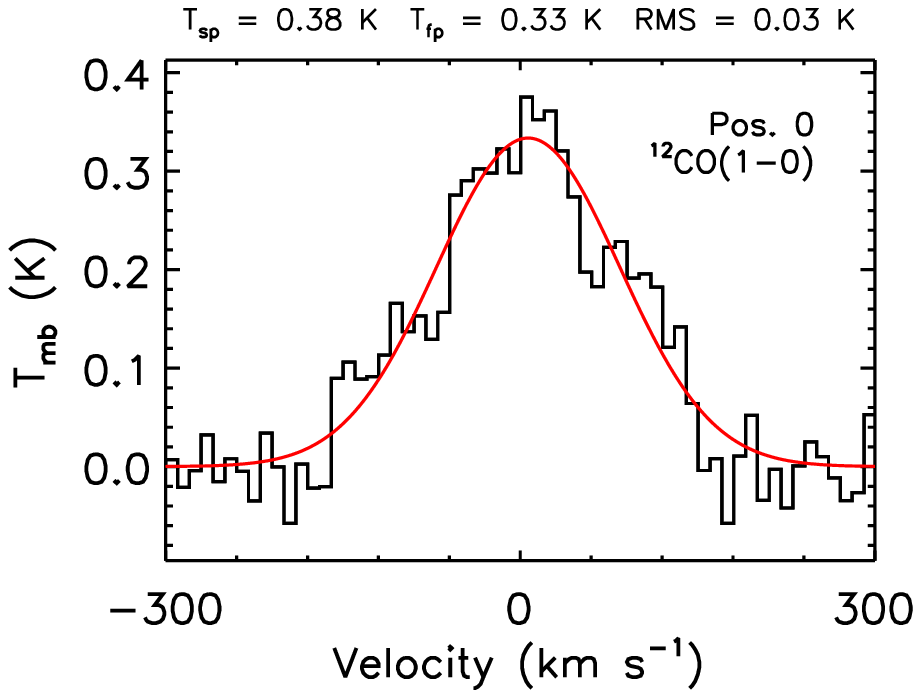}
  \includegraphics[width=5.8cm,clip=]{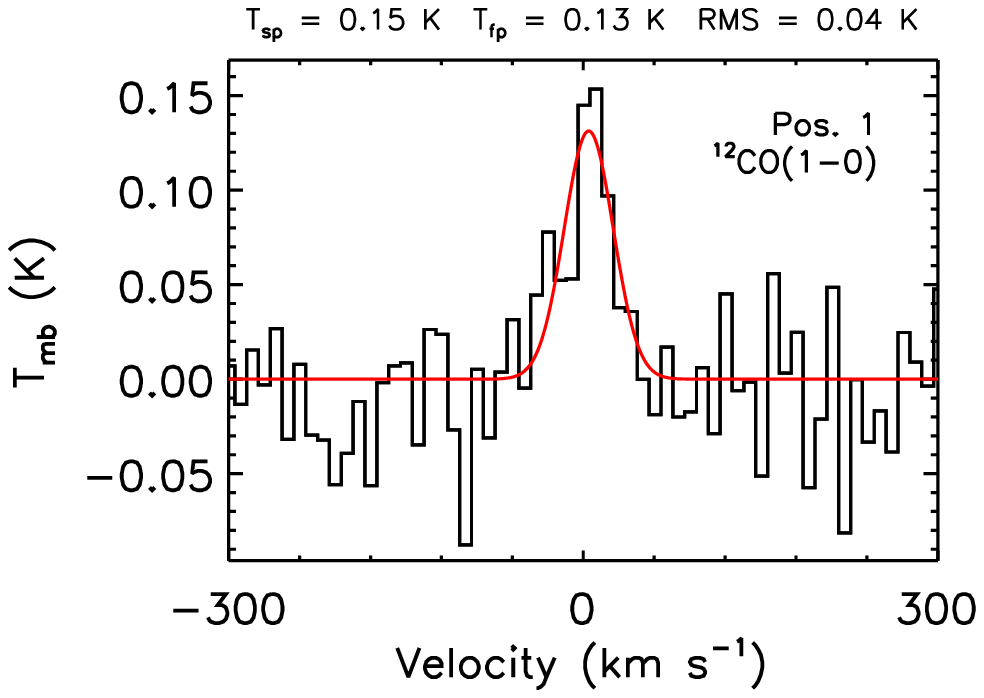}
    \includegraphics[width=5.8cm,clip=]{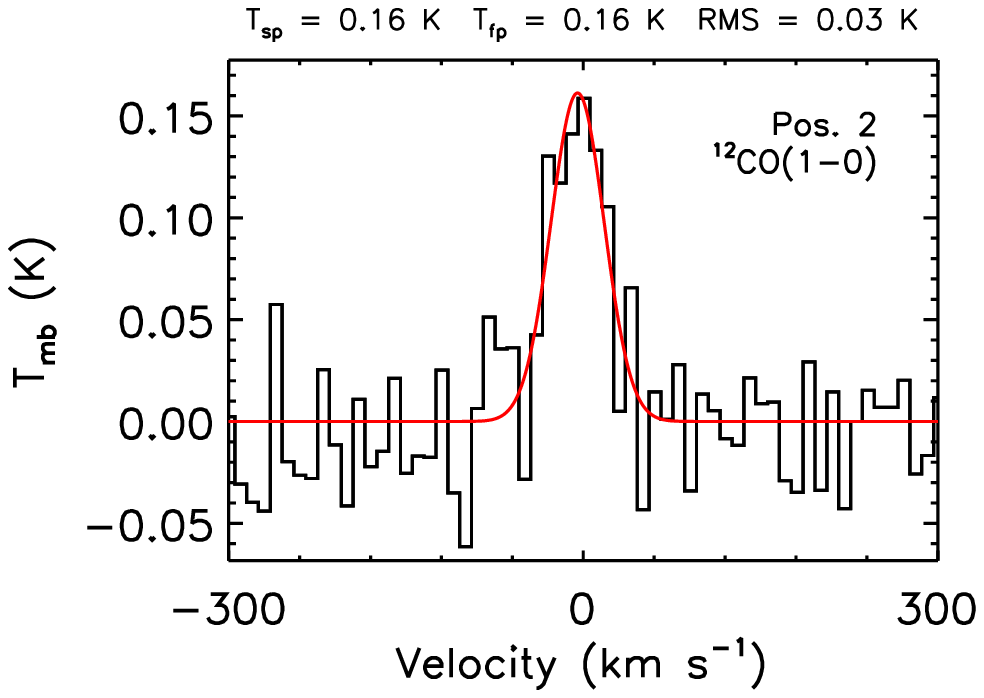}\\
  \vspace{-5pt}
  \hspace{-15pt}
  \includegraphics[width=5.8cm,clip=]{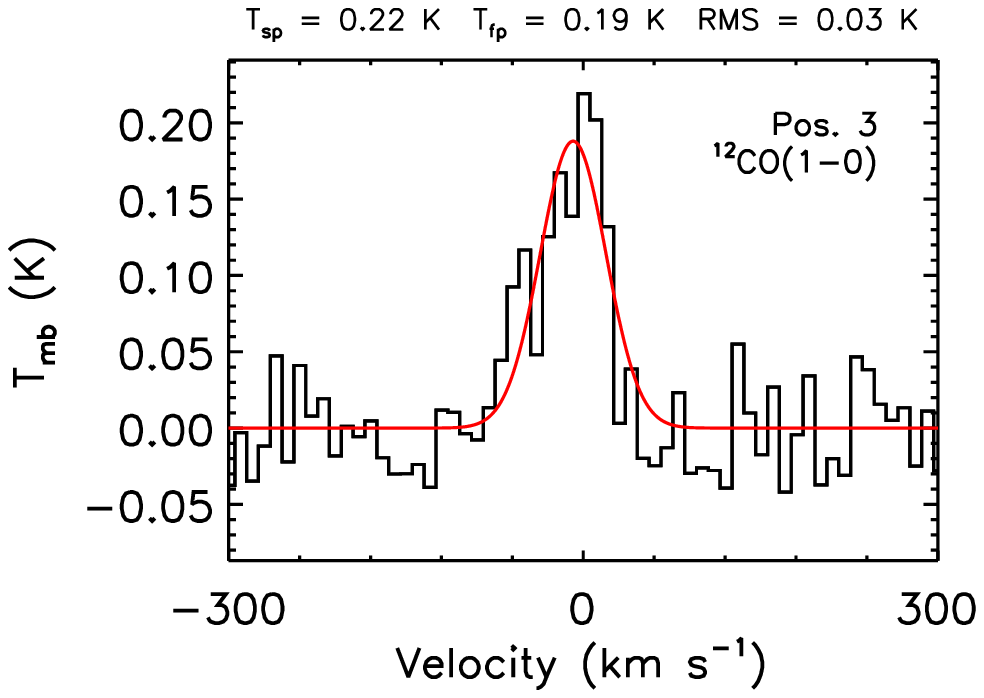}
  \includegraphics[width=5.8cm,clip=]{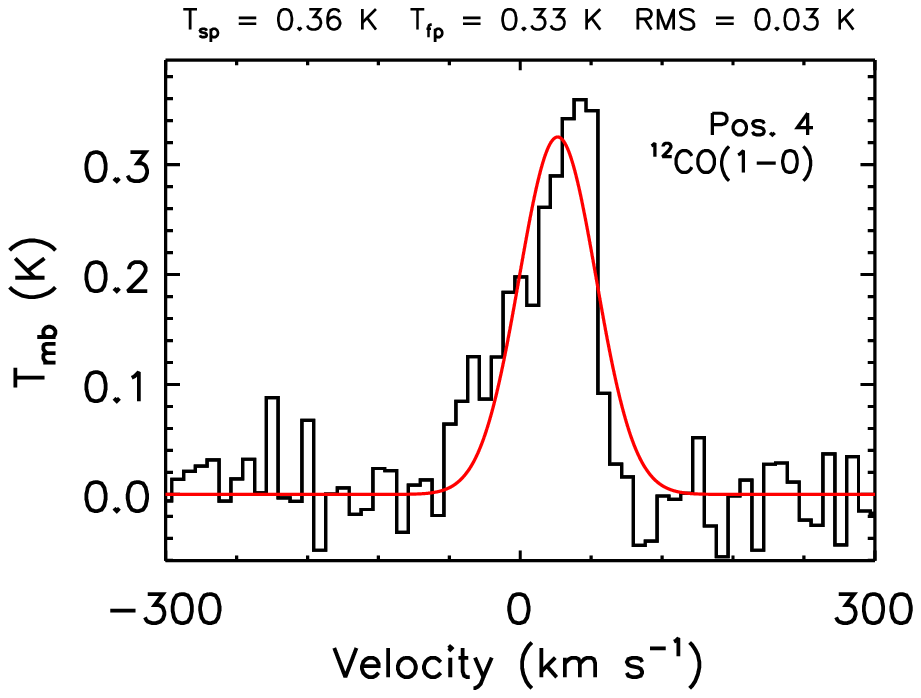}
    \includegraphics[width=5.8cm,clip=]{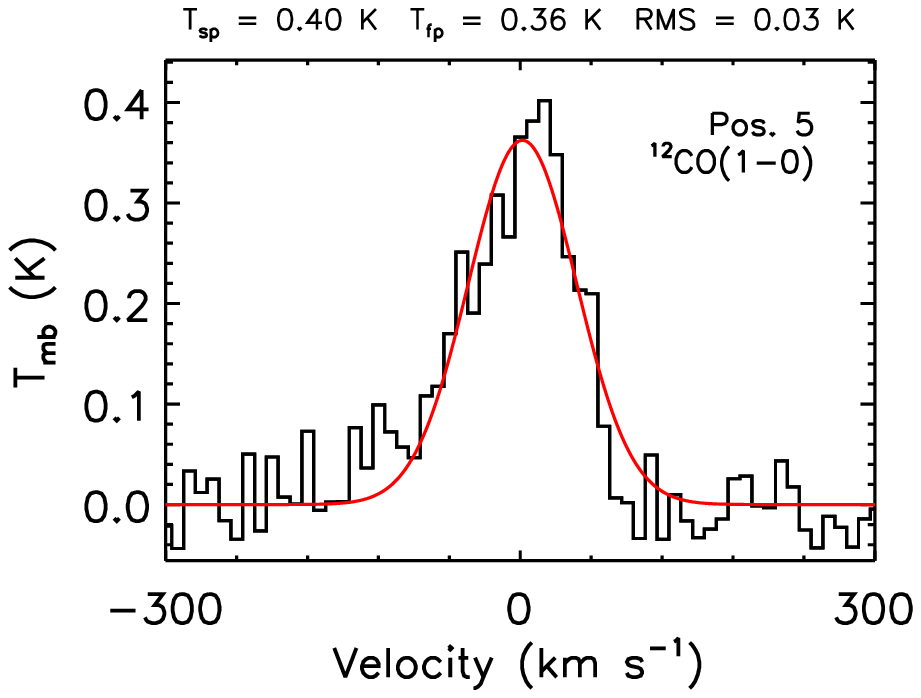}\\
  \vspace{-5pt}
  \hspace{-15pt}
  \includegraphics[width=5.8cm,clip=]{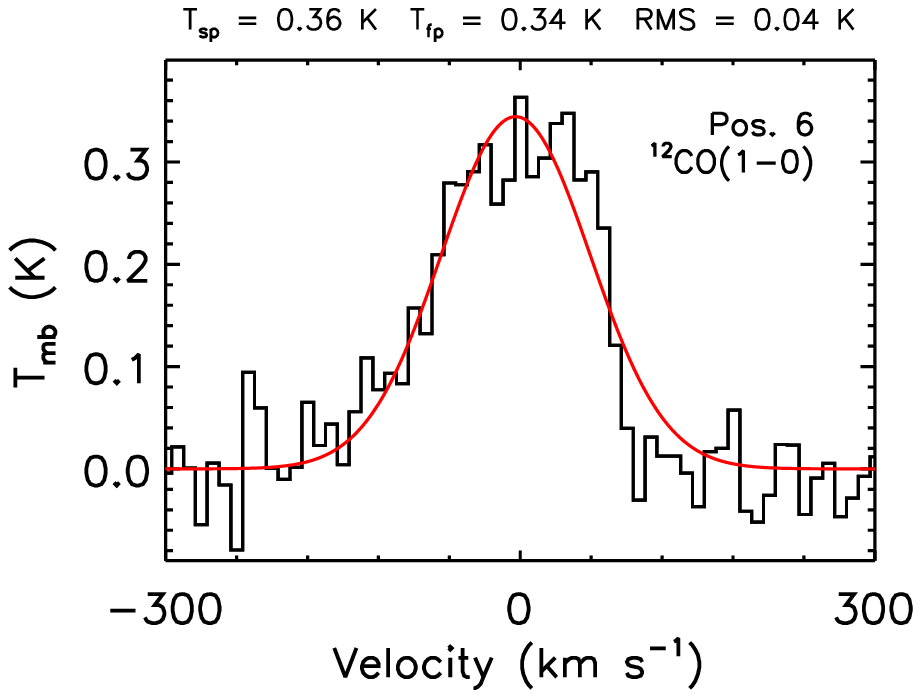}
  \includegraphics[width=5.8cm,clip=]{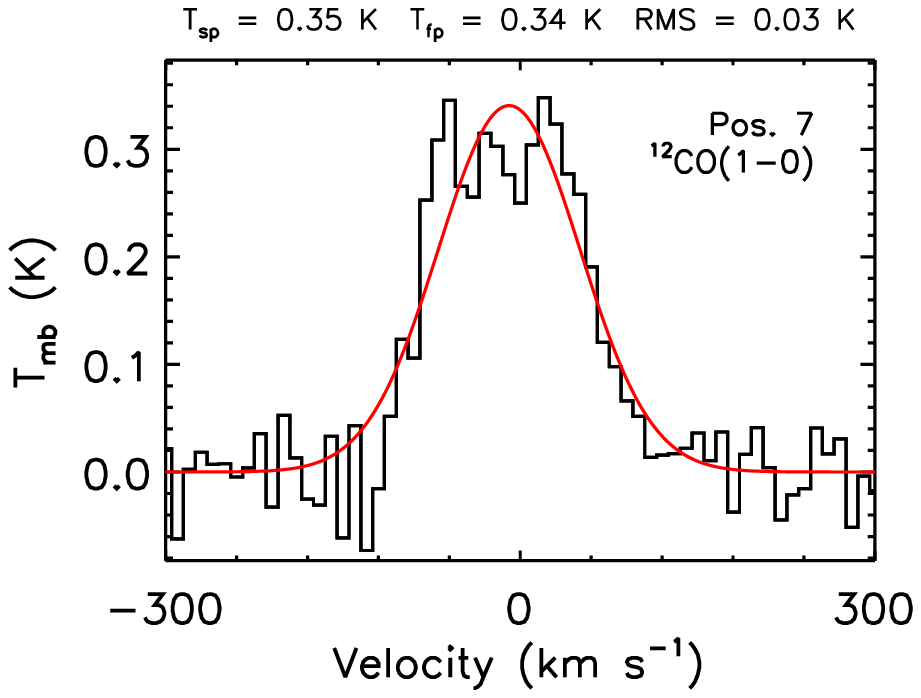}
    \includegraphics[width=5.8cm,clip=]{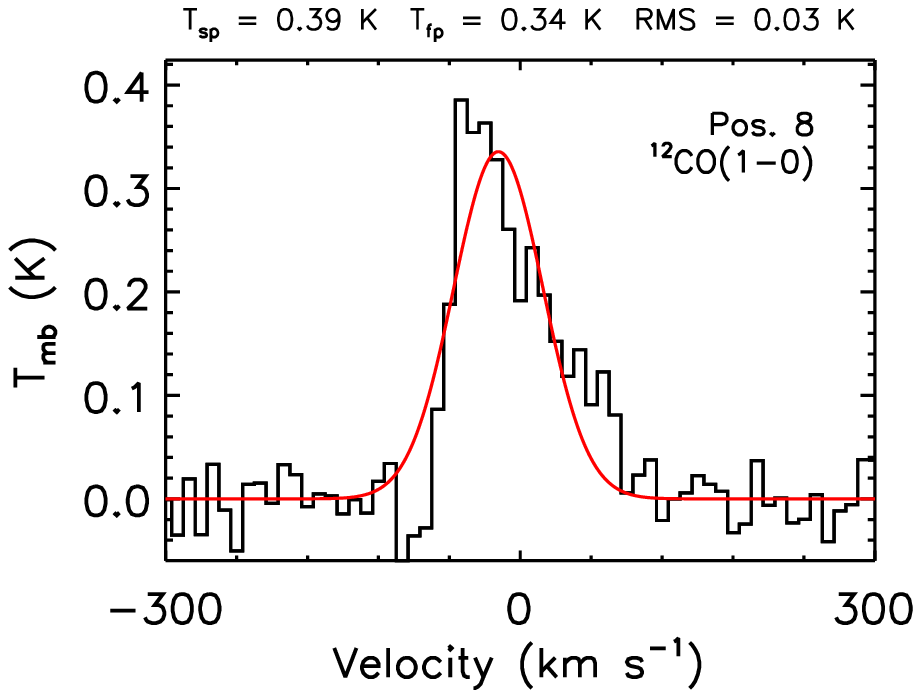}\\
      \vspace{-5pt}
  \hspace{-15pt}
  \includegraphics[width=5.8cm,clip=]{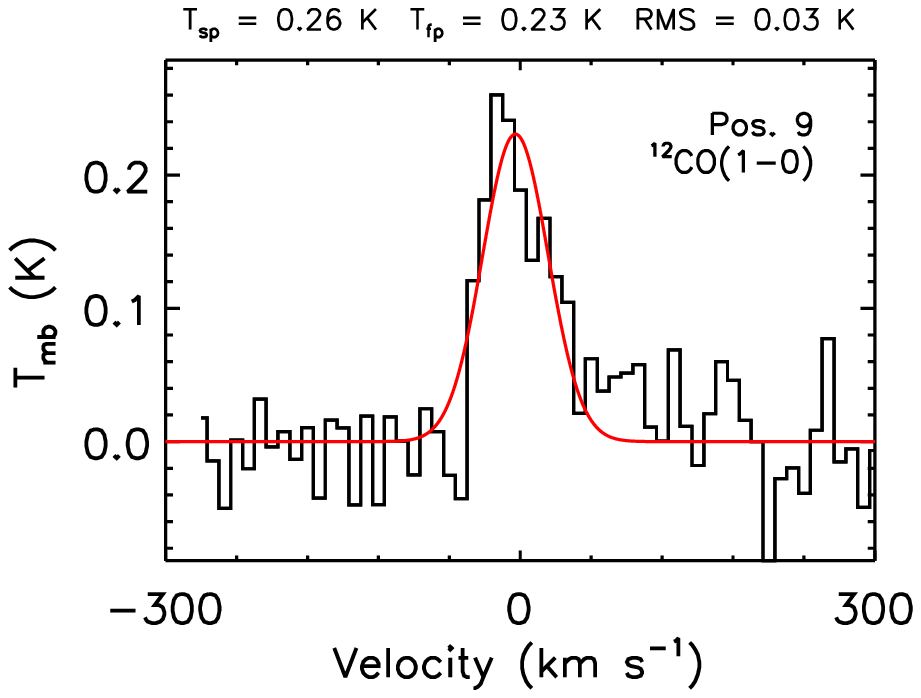}
  \includegraphics[width=5.8cm,clip=]{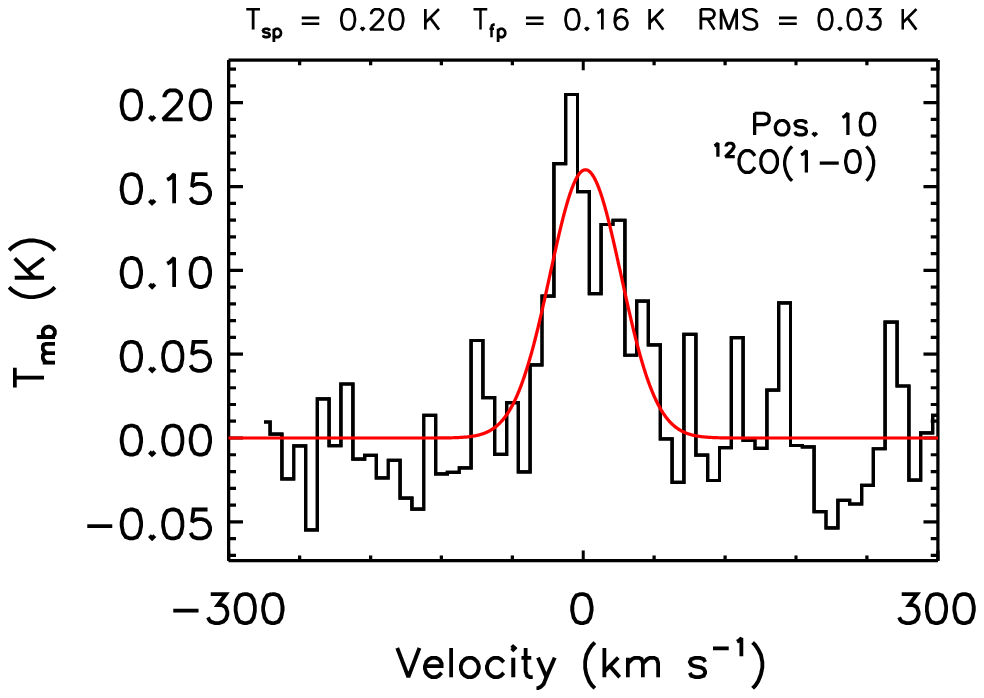}
    \includegraphics[width=5.8cm,clip=]{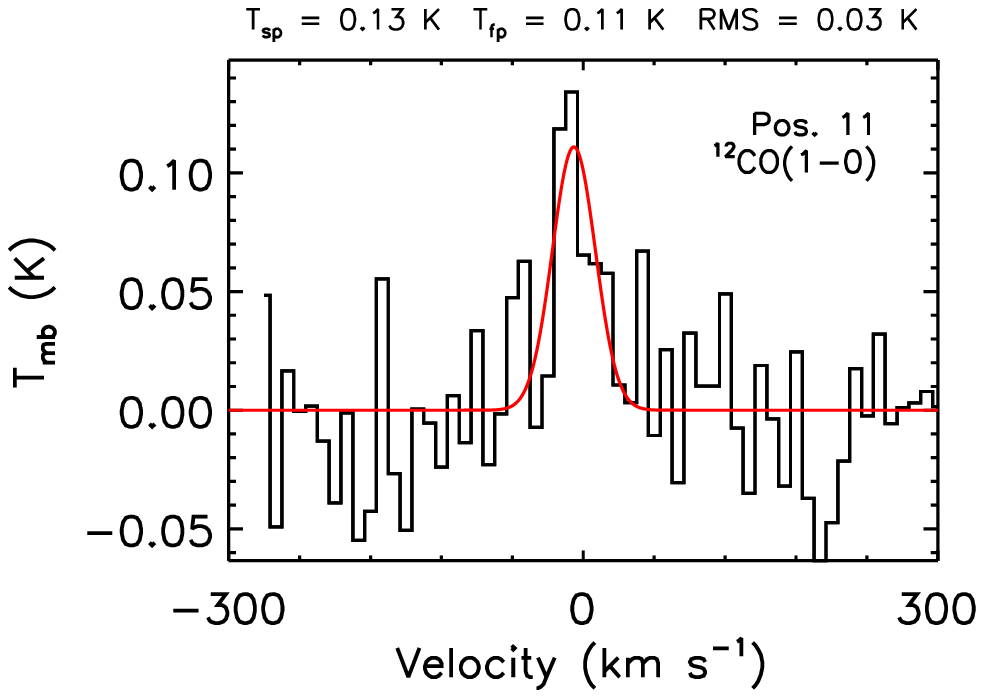}\\
      \vspace{-5pt}
  \hspace{-15pt}
    \includegraphics[width=5.8cm,clip=]{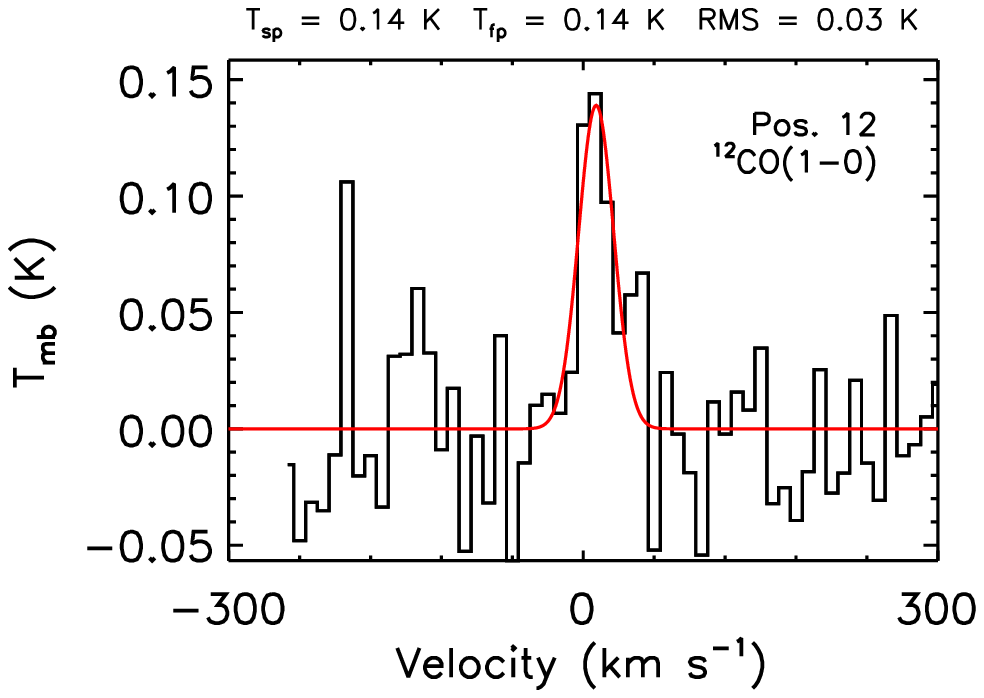}
      \caption{Integrated $^{12}$CO(1--0) profiles extracted from the positions over the NGC~1055 are shown with the best-fit Gaussian function (red line). The peak brightness temperature for the spectra, i.e. T$_{\rm sp}$, and for the fit, i.e. T$_{\rm fp}$, are also shown over each spectrum. The RMS represents the standard deviation of the brightness temperatures in the emission-free part of the spectra.}
  \label{fig:profit}
\end{figure*}

\addtocounter{figure}{-1}
\begin{figure*}
  \hspace{-15pt}
  \includegraphics[width=5.8cm,clip=]{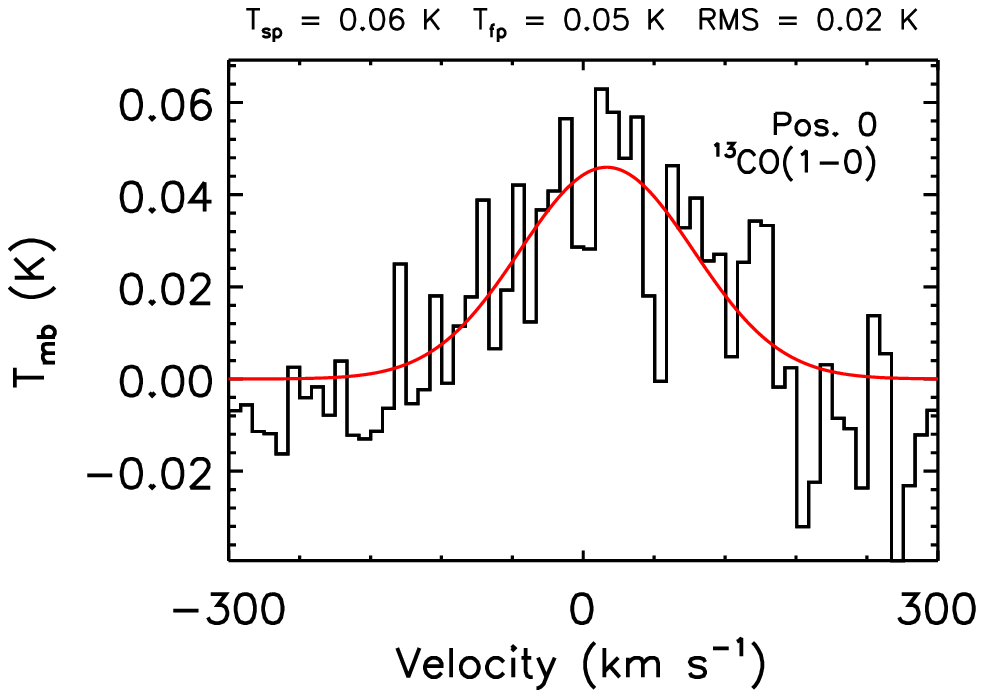}
  \includegraphics[width=5.8cm,clip=]{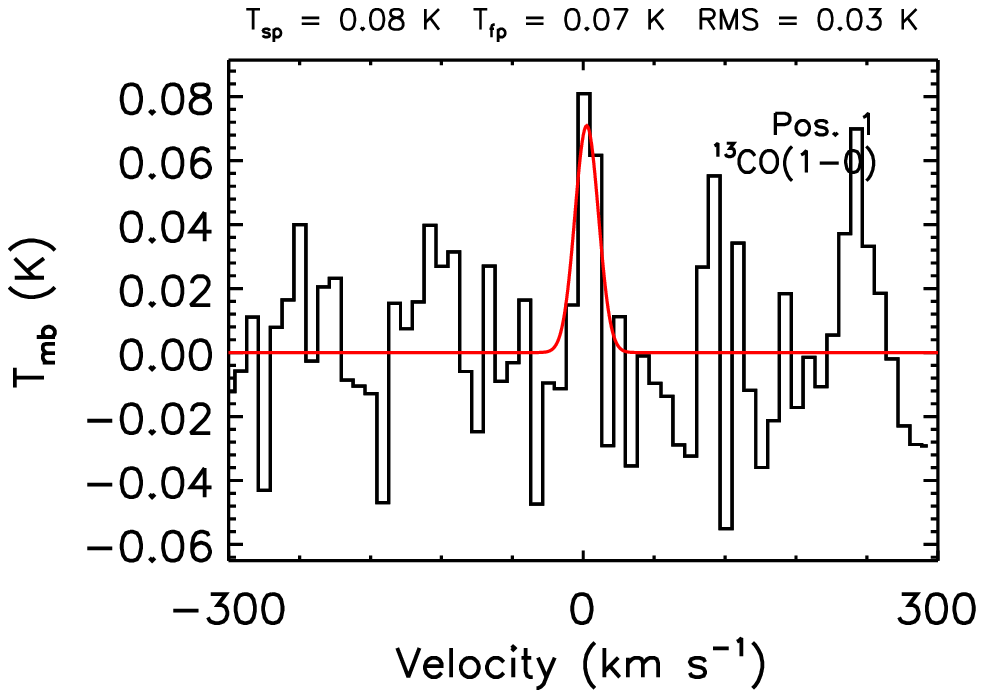}
    \includegraphics[width=5.8cm,clip=]{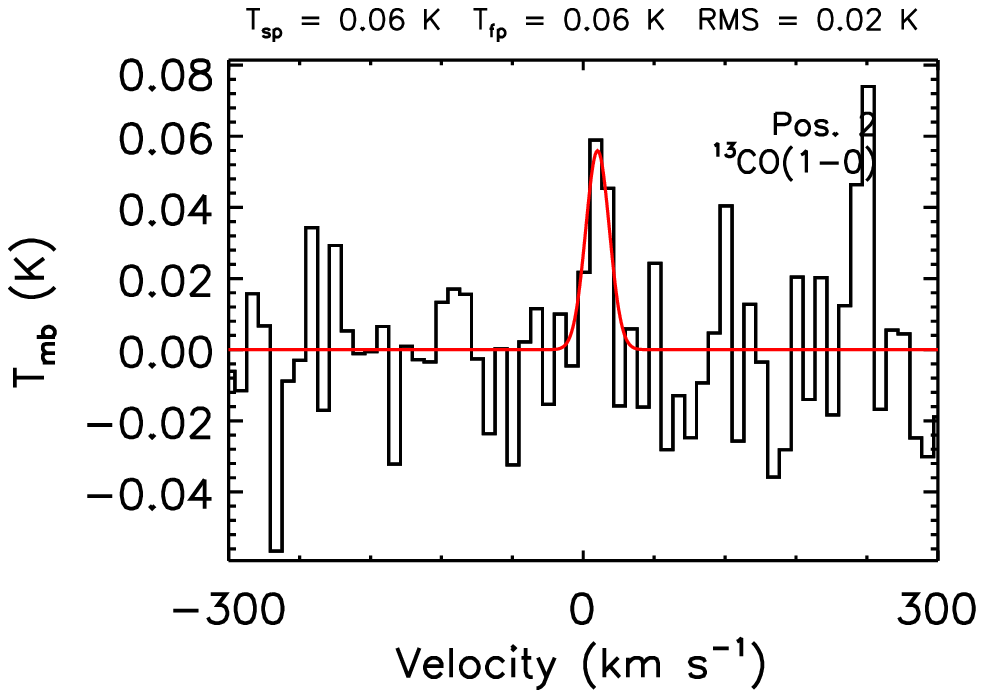}\\
  \vspace{-5pt}
  \hspace{-15pt}
  \includegraphics[width=5.8cm,clip=]{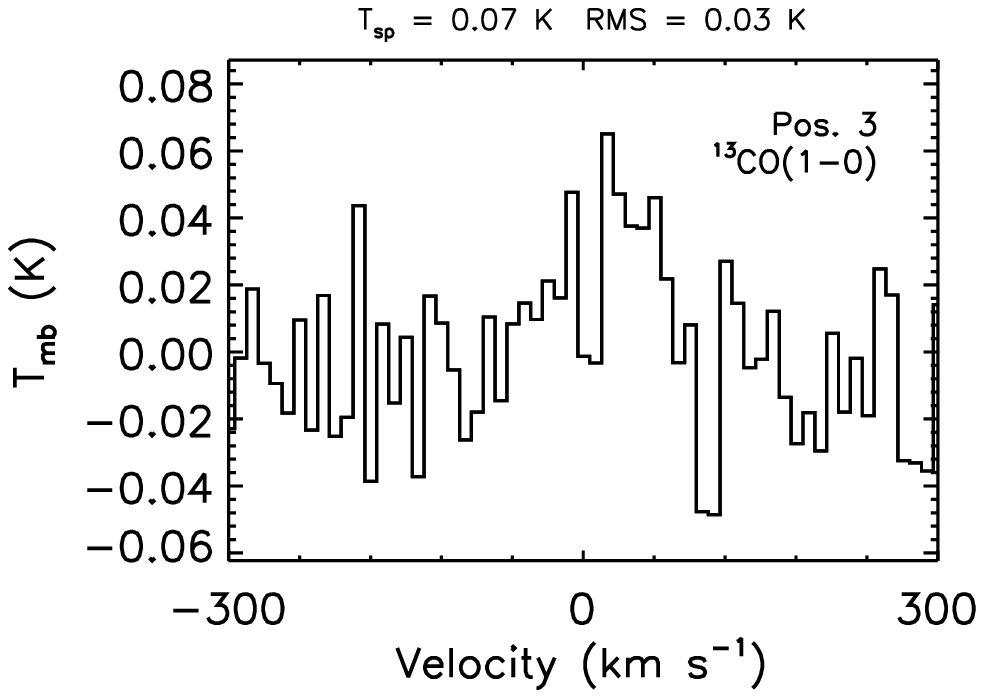}
  \includegraphics[width=5.8cm,clip=]{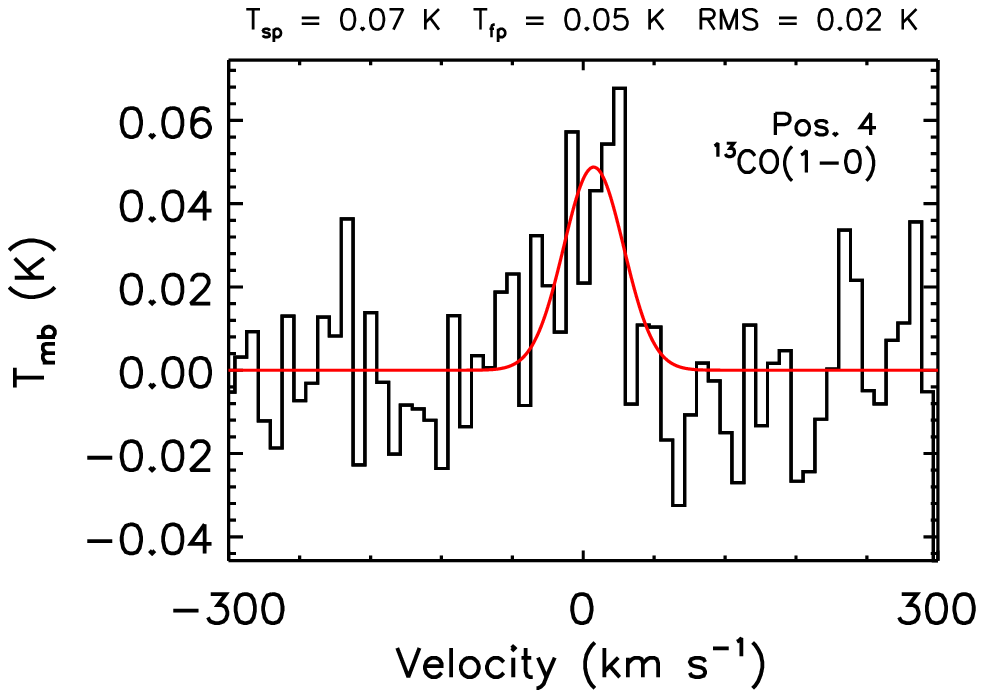}
    \includegraphics[width=5.8cm,clip=]{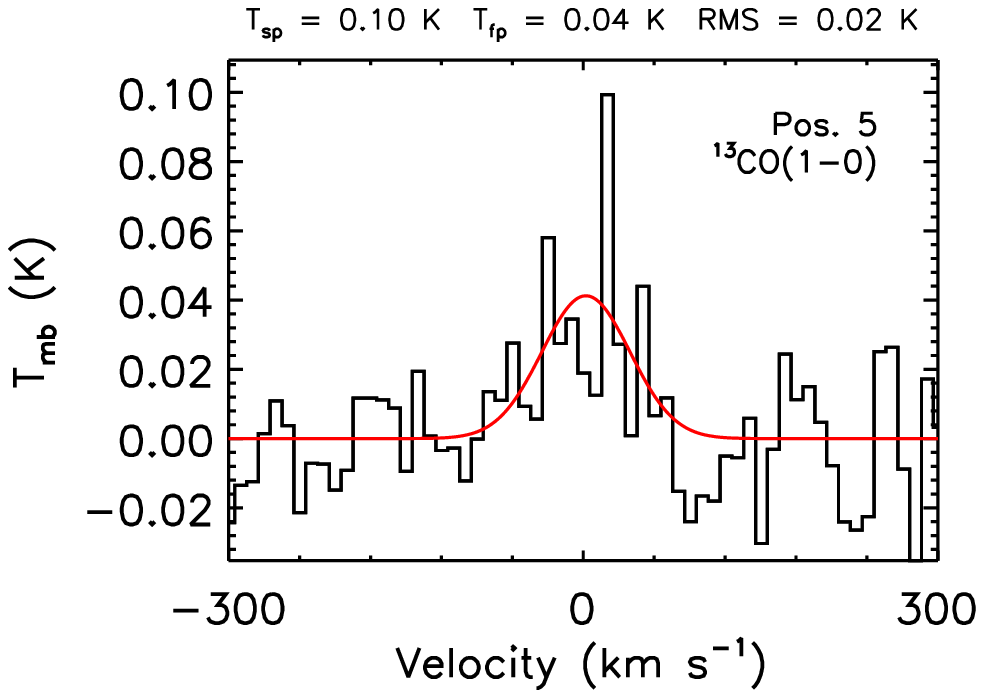}\\
  \vspace{-5pt}
  \hspace{-15pt}
  \includegraphics[width=5.8cm,clip=]{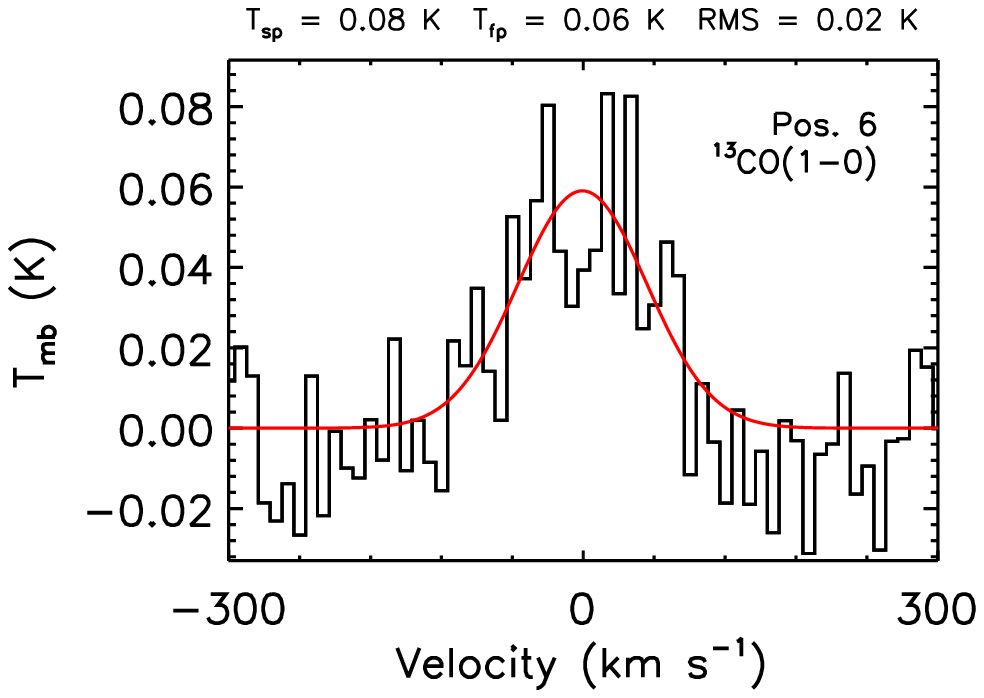}
  \includegraphics[width=5.8cm,clip=]{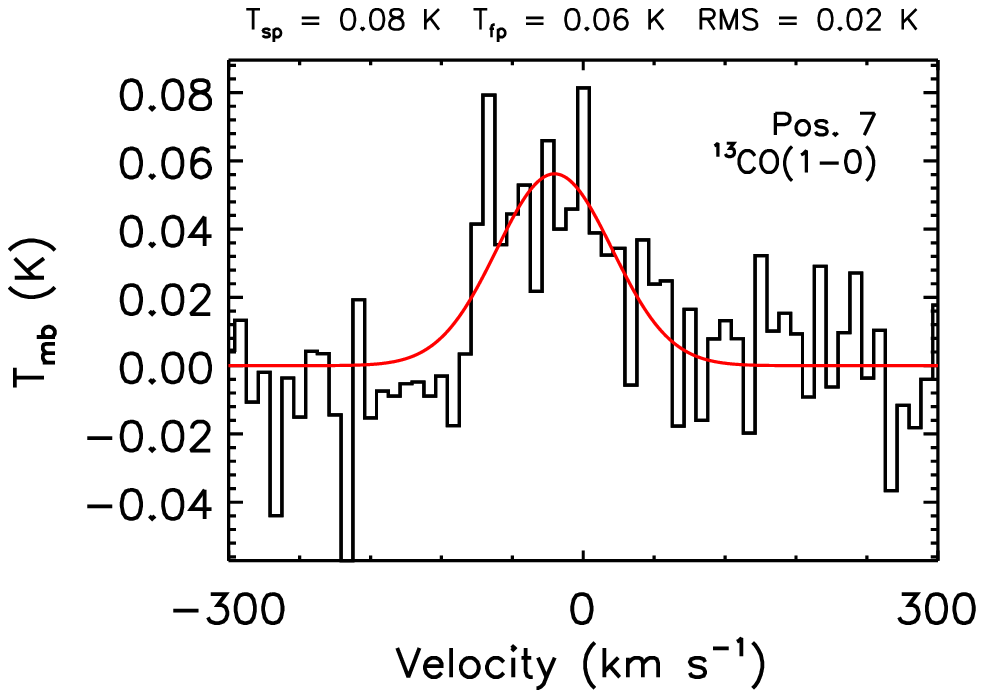}
    \includegraphics[width=5.8cm,clip=]{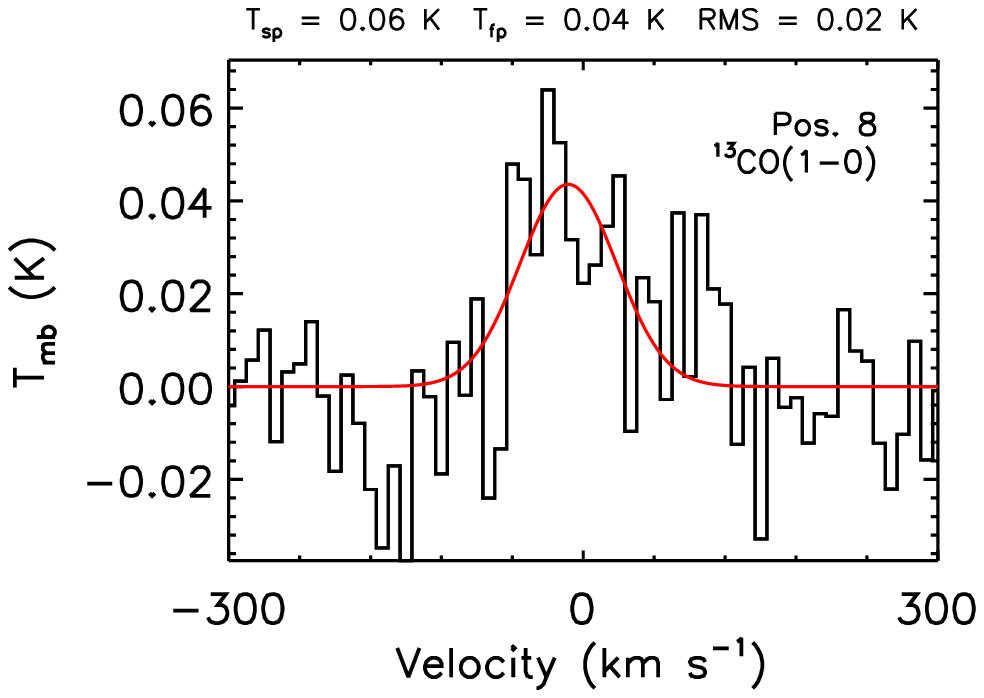}\\
      \vspace{-5pt}
  \hspace{-15pt}
  \includegraphics[width=5.8cm,clip=]{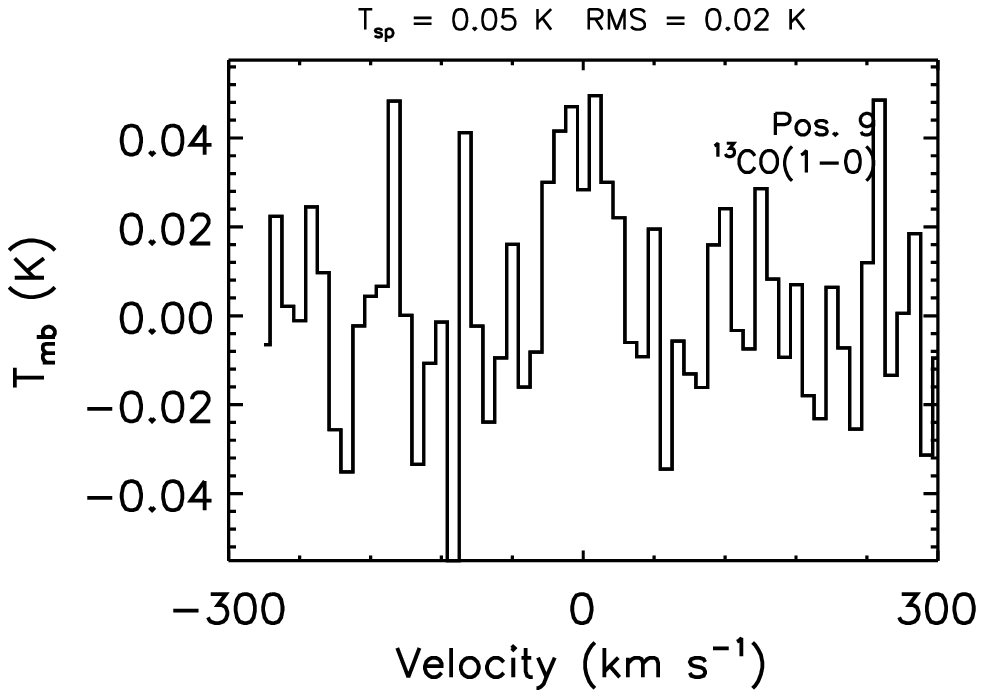}
  \includegraphics[width=5.8cm,clip=]{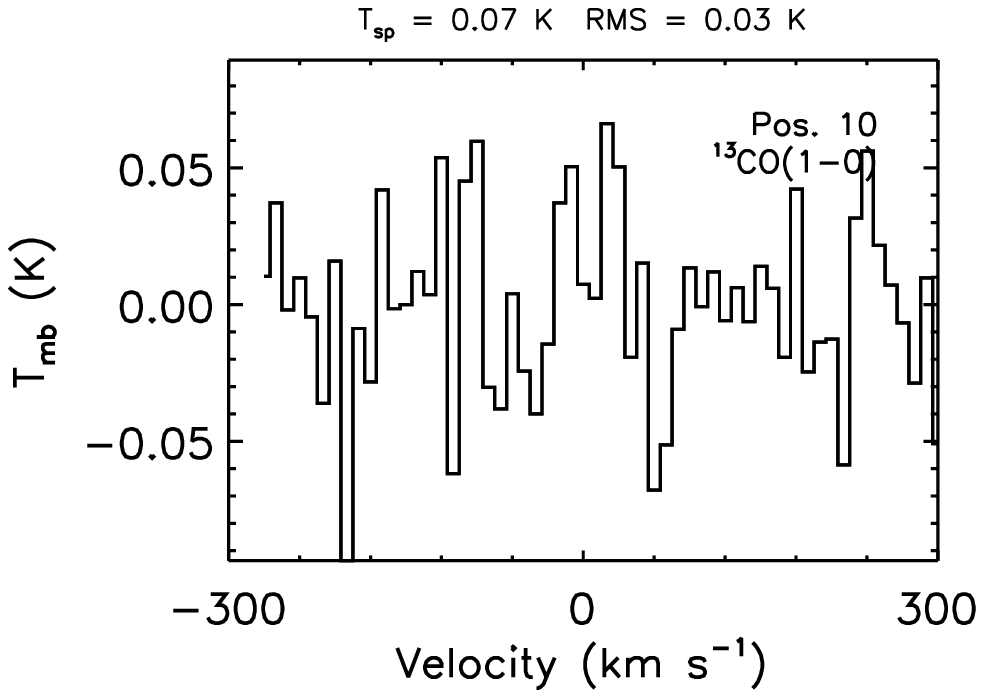}\\
      \caption{Continued but for $^{13}$CO(1--0). The positions with non-detection (i.e. $\le\,2.5\sigma$) do not include the best fit shown by the red line and the associated peak temperature, i.e. T$_{\rm fp}$.}
  \label{fig:profit}
\end{figure*}

\end{document}